\documentclass[aps,pre,amsmath,twocolumn,superscriptaddress,showpacs]{revtex4-2}
\usepackage{calc}
\usepackage{amsmath}
\usepackage{amsfonts}
\usepackage{amssymb}
\usepackage{graphicx}
\usepackage{hyperref}
\usepackage{soul}
\usepackage{mathtools}
\usepackage{verbatim}
\usepackage{epstopdf}
\usepackage{subfigure}
\usepackage{latexsym}
\usepackage{dcolumn}
\usepackage{epsf}
\usepackage{float}
\usepackage[table]{xcolor}
\usepackage{multirow}
\usepackage[toc]{appendix}

\usepackage{color} 

\usepackage{graphicx}
\usepackage{graphicx,epstopdf}
\usepackage{xcolor}
\usepackage{dcolumn}
\usepackage{bm}
\usepackage{mathrsfs} 

\usepackage{mathrsfs}
\begin{document}

\title{Periodic Environmental Forcing Shapes the Stability of Complex Ecological Networks} 

\author{Sayantan Nag Chowdhury}
\email{jcjeetchowdhury1@gmail.com}

\affiliation{Department of Mathematics, Nalanda University, Rajgir, Bihar 803116, India}

\date{\today}

\begin{abstract}


Environmental variability is a defining feature of natural ecosystems, yet many theories of ecological stability assume static environments. {\color{black}Here, we develop an analytical framework for local stability in complex ecological communities subjected to periodic environmental forcing. We show that, in slowly varying environments and under a large-system approximation, the maximal Lyapunov exponent can be approximated by the time-averaged rightmost spectral edge of the instantaneous interaction matrix, yielding explicit stability criteria for large ecological communities. Within the random-matrix interaction ensembles considered here, our minimal theoretical framework predicts a topology-dependent ordering of local perturbation stability under periodic temporal modulation in high-dimensional ecological networks, which is supported by numerical simulations.} Beyond the adiabatic regime, rapid environmental oscillations dynamically stabilize otherwise unstable ecosystems, revealing a high-frequency rescue effect that is absent from static theories. These results extend ecological stability theory beyond autonomous systems and provide a general framework for understanding resilience in fluctuating environments.

\end{abstract}

\maketitle

\section{Introduction}

 One of the most enduring paradoxes in theoretical ecology is the persistent survival of highly diverse, densely connected ecosystems \cite{artime2024robustness,landi2018complexity}. Natural ecosystems are perhaps among the most complex dynamical systems found in nature. Thousands of species interact through predation, competition, mutualism, and other complex ecological relationships, forming intricate networks that continuously evolve in time. Despite this complexity, ecological communities often persist for remarkably long periods, maintaining their overall structure in the face of environmental disturbances, seasonal fluctuations, and demographic variability. Understanding the mechanisms that enable complex ecosystems to remain stable has therefore been a central objective of theoretical ecology for more than half a century \cite{sanders2024ecosystem,barabas2017self,rooney2012integrating,montoya2006ecological,meena2023emergent}.

The modern study of ecological stability \cite{dominguez2019unveiling,chen2024stability} was profoundly influenced by the pioneering work of Robert May \cite{may1972will}, who demonstrated more than half a century ago that increasing complexity does not necessarily promote stability in large ecological communities. Using random matrix theory, May showed that ecosystems characterized by strong interactions, high species richness, or dense connectivity can become unstable beyond a critical threshold.  Yet, looking at the natural world, from teeming coral reefs to dense tropical rainforests, complexity is the rule, not the exception. This discovery initiated a long-standing debate \cite{mccann2000diversity} regarding the relationship between complexity and stability and motivated extensive efforts to identify structural features that enhance ecological resilience. Over the past decade, an important piece of this puzzle has emerged from the realization that natural ecosystems are not wired at random. Species interact through highly structured, biologically constrained relationships. Random matrix approaches have shown that the correlation between reciprocal interactions, whether species interact as predators and prey, competitors, or mutualists, fundamentally alters community stability. In particular, antagonistic predator-prey interactions tend to suppress instability, whereas symmetric competitive and mutualistic interactions amplify perturbations and reduce resilience \cite{allesina2012stability,mougi2012diversity,allesina2008network}.

While these advances have significantly improved our understanding of ecological stability, an even deeper question remains unanswered: {\it how do vast ecological networks maintain their delicate balance as the environment itself is in a constant state of flux?} Real ecosystems do not exist in a static environment; they are constantly influenced by environmental fluctuations \cite{ripa2003food,vasseur2005mechanistic} ranging from temperature swings to seasonal cycles. Despite the advances, most theoretical studies of ecological stability remain confined to autonomous dynamical systems, in which most frameworks explicitly assume a perfectly static environment. In reality, however, natural ecosystems are continuously exposed to environmental fluctuations operating across a wide range of temporal scales. Seasonal cycles, climate oscillations, and other recurrent environmental processes can modify interaction strengths, alter demographic rates, and reshape community structure. A severe drought or a sudden heatwave can fundamentally alter the rates at which species consume, compete, or cooperate. Thus, environmental forcing continuously modulates the strength of ecological interactions. Consequently, ecological interactions are inherently time-dependent, and ecosystem stability should be viewed as a dynamical property that emerges from the interplay between network structure and environmental forcing.

Incorporating explicit temporal variability into ecological stability theory presents a major theoretical challenge. Time-dependent ecological networks generally cannot be characterized by the eigenvalue spectrum of a single static interaction matrix. Instead, stability must be quantified using dynamical measures such as Lyapunov exponents; thus, investigating the stability of such non-autonomous \cite{kloeden2011nonautonomous, caraballo2017applied}, time-varying networks \cite{holme2012temporal,ghosh2022synchronized,berner2023adaptive} has required computationally expensive numerical integration. While brute-force simulations can map the boundaries of survival and extinction, they often act as a black box, obscuring the underlying mathematical mechanisms that govern why an ecosystem survives a fluctuating environment. As a result, despite growing interest in environmentally driven ecological dynamics, a predictive analytical framework for understanding the stability of large, time-dependent ecosystems remains largely absent.

In this study, we address this challenge by investigating the stability of ecological communities subjected to periodic environmental forcing. We deliberately adopt a minimal mathematical framework. Although real ecosystems exhibit nonlinear population dynamics, multiple environmental drivers, species-specific responses, and interactions that vary across space and time, our goal is not to incorporate every ecological detail \cite{kot2001elements,hastings2013population,brauer2012mathematical,rockwood2015introduction,inaba2017age}. Rather, we seek to identify the minimal ingredients necessary to understand how environmental variability influences the stability of complex communities. To this end, we focus on a linearized description of population dynamics near equilibrium, for which stability is naturally quantified by the maximal Lyapunov exponent.

{\color{black}The present framework is deliberately formulated at the level of local stability rather than full nonlinear population dynamics. The state vector represents an infinitesimal perturbation about a reference equilibrium, and the maximal Lyapunov exponent therefore quantifies the asymptotic growth or decay of perturbations. The model does not explicitly represent resource
depletion, species-specific demographic rates, carrying capacities, or nonlinear functional responses. Accordingly, the stability results should not be interpreted as direct predictions of species persistence, extinction,
or population abundance. Instead, the aim is to isolate how interaction topology and periodic temporal modulation affect local stability in a high-dimensional community-matrix framework.}

We represent the fluctuating ecosystem using two interaction matrices: a baseline community matrix describing the intrinsic topological structure of the ecological network, and an environmental coupling matrix that characterizes how external fluctuations modify interaction strengths. This separation allows environmental forcing to alter the intensity of ecological interactions without changing the underlying network architecture. Ecologically, this reflects the fact that environmental change often modifies the strength of interactions rather than their fundamental nature. Mathematically, it provides a transparent framework for distinguishing the intrinsic properties of the ecosystem from the influence of the external environment.

The environmental forcing is assumed to be periodic. This choice serves as the simplest representation of recurring environmental variability, including seasonal cycles, climatic oscillations, and periodic resource availability. Far from being merely a mathematical convenience, periodic forcing captures some of the most ubiquitous temporal patterns observed in nature. Moreover, the chosen periodic forcing constitutes the fundamental Fourier building block of arbitrary periodic signals, making it a natural starting point for understanding more complex environmental fluctuations. As we demonstrate below, this deliberately simple yet biologically motivated framework already gives rise to rich and nontrivial behavior, including topology-dependent stability, analytically predictable stability boundaries, and a high-frequency stabilization mechanism.

Using random matrix theory and Lyapunov exponent analysis, we investigate the stability of three canonical ecological architectures: random, predator-prey, and competition-mutualism community. Exploiting the large-system limit and the elliptic law \cite{girko1985circular} for random matrices with correlated interactions, we derive an analytical approximation for the maximal Lyapunov exponent in the regime of slowly varying environmental forcing. The resulting theory predicts the maximal Lyapunov exponent from the time-averaged rightmost spectral edge of the instantaneous community matrix and yields explicit stability boundaries in terms of interaction strength, connectance, self-regulation, forcing amplitude, and interaction topology. {\color{black}Within the random-matrix ensembles considered in this study, we} find that predator-prey communities are most stable, competition-mutualism is the least stable, and random communities fall in between. Simulations show that rapid environmental oscillations suppress perturbation growth through dynamical averaging \cite{,sanders2007averaging}, thereby expanding the stable parameter region. These results create a unified theory connecting interaction topology, environmental forcing, and stability, and demonstrate how temporal variability can reshape ecosystem resilience and the importance of incorporating explicit environmental dynamics into ecological stability theory.

\section{Mathematical Model}

To explore how periodic environmental forcing influences the stability of complex ecological networks, we develop a minimal time-dependent linearized framework and examine three biologically motivated interaction architectures \cite{allesina2012stability}: random, predator-prey, and competition-mutualism communities.
 
\subsection{Theoretical Framework}
The local stability of an ecological community near an equilibrium state can be assessed by analyzing the evolution of small perturbations in species abundances. In a static environment, the dynamics of a perturbation vector $\mathbf{x}(t)$ for an ecosystem consisting of $S$ interacting species are governed by the linearized equation
\[
\frac{d\mathbf{x}}{dt} = (-dI+B)\mathbf{x},
\]

{\color{black} where $d$ denotes the common linear self-regulation (intrinsic damping) rate, $I$ is the $S\times S$ identity matrix, and $B$ is the baseline community interaction matrix. Because $x(t)$ represents an infinitesimal perturbation rather than the absolute abundance vector, the parameter $d$ should be interpreted here as a linear damping or self-regulation scale rather than as a species-specific difference between birth and death rates.} The balance between self-regulation and interspecific interactions determines whether perturbations decay or grow over time \cite{may1972will,allesina2012stability}.


Real ecosystems, however, are rarely exposed to static environmental conditions. Seasonal cycles, climatic oscillations, and recurrent environmental disturbances continuously modify interaction strengths and alter community dynamics. To account for such effects, we consider a periodically forced ecological network described by

\begin{equation}
\frac{d\mathbf{x}}{dt} = A(t) \mathbf{x}= \left[ -dI + B + \varepsilon \cos(\omega t)H \right] \mathbf{x},
\label{eq:main_dynamics}
\end{equation}
where $\mathbf{x}(t)\in\mathbb{R}^{S}$ represents an infinitesimal state perturbation away from a feasible equilibrium of the ecosystem \cite{gross2009generalized,may2001stability}.
The matrix $B$ encodes the baseline ecological interactions, while the matrix $H$ characterizes the sensitivity of those interactions to environmental forcing. The parameter $\varepsilon$ controls the amplitude of the forcing and $\omega$ denotes its frequency. The term $(-dI + B)\mathbf{x}$ captures the intrinsic transient dynamics of the state perturbation in a static environment. The term $\left[ \varepsilon \cos(\omega t)H \right] \mathbf{x}$ represents the bilinear coupling between the state perturbation and the environmental forcing (i.e., it is linear with respect to both the perturbation vector $\mathbf{x}$ and the forcing amplitude $\varepsilon$); it mathematically dictates how the fluctuating environment continuously reshapes the recovery or collapse trajectory of the displaced populations. {\color{black}For fixed forcing parameters, Eq.\ \eqref{eq:main_dynamics} remains linear and homogeneous in the perturbation vector $x$, while being nonautonomous because its coefficient matrix varies periodically in time.}

The environmental susceptibility matrix $H$ is generated independently of $B$. This assumption reflects the idea that the ecological interactions and their responses to environmental variability need not be directly correlated. Consequently, $H$ specifies how strongly a given interaction is amplified or weakened by external forcing, independently of its baseline strength in $B$. The self-regulation term $-dI$ introduces stabilizing feedback that suppresses unbounded population growth, whereas periodic environmental variability is incorporated through the oscillatory contribution $\varepsilon\cos(\omega t)H$, which continuously modulates interaction strengths in time. 

\subsection{Interaction Architectures}

To investigate the role of interaction topology in determining ecological stability, we consider three distinct network architectures \cite{allesina2012stability}: random, predator-prey, and competition-mutualism community. In all cases, both the baseline interaction matrix $B$ and the environmental susceptibility matrix $H$ are constructed as sparse matrices characterized by a connectance parameter $C$, which specifies the probability that any pair of species interacts. The magnitude of each non-zero interaction is determined by the parameter $\sigma$, which controls the overall interaction strength.

\paragraph{random communities.}
For random communities, the off-diagonal elements of both $B$ and $H$ are independently drawn from a Gaussian distribution,
$$B_{ij}, H_{ij} \sim \mathcal{N} \left( 0, \frac{\sigma^2}{S} \right), \qquad i\neq j.$$
Network sparsity is imposed through connectance $C$, such that each interaction is retained with probability $C$ and removed otherwise. Consequently, the matrices possess no prescribed sign structure and serve as a neutral reference model.

Note that the interaction coefficients are drawn from a distribution with variance $\frac{\sigma^2}{S}$. This normalization is standard in random matrix theory and is essential for obtaining a well-defined large-system limit, ensuring that the total ecological feedback remains finite as the community size $S \to \infty$ increases. From an ecological perspective, this scaling is consistent with the assumption that, as species richness increases, the average strength of individual pairwise interactions decreases, thereby preventing the cumulative interaction pressure from diverging and maintaining biologically meaningful community dynamics.

\paragraph{predator-prey communities.}
For predator-prey communities, interacting species pairs are assigned opposite signs, reflecting antagonistic trophic interactions. Specifically,
$$(B_{ij},B_{ji}) = (+g_1,-g_2) \quad\text{or}\quad (-g_1,+g_2),$$
where
$$g_1, g_2 \sim \left| \mathcal{N} \left( 0, \frac{\sigma^2}{S} \right) \right|.$$
Thus, whenever one species benefits from an interaction, the other species is negatively affected.

\paragraph{competition-mutualism communities.}
For competition-mutualism communities, interacting species pairs share the same sign. Thus,
$$(B_{ij},B_{ji}) = (+g_1,+g_2)$$
represents a mutualistic interaction, while
$$(B_{ij},B_{ji}) = (-g_1,-g_2)$$
corresponds to a competitive interaction, where
$$g_1, g_2 \sim \left| \mathcal{N} \left( 0, \frac{\sigma^2}{S} \right) \right|.$$


{\color{black} The matrices $B$ and $H$ are independently sampled from the same statistical interaction ensemble (random, predator-prey, or competition-mutualism), with the same connectance $C$, interaction class, and conditional distribution of nonzero entries. They do not, however, share the same realization of the connectivity pattern. Thus, $B$ and $H$ have the same statistical interaction architecture while representing independent realizations of the baseline interaction structure and environmental susceptibility.} This choice reflects the assumption that environmental forcing modulates the strength of ecological interactions without altering their fundamental topological character. Ecologically, environmental fluctuations typically modify the intensity or efficiency of existing interactions rather than transforming one interaction type into another. For example, a drought may increase or decrease the efficiency with which a predator captures its prey, but it does not fundamentally convert a predator-prey relationship into a mutualistic one. Consequently, the environmental coupling matrix $H$ is assumed to preserve the same statistical interaction structure as the baseline community matrix $B$. From a modeling perspective, this assumption also enables a clean separation between two distinct mechanisms: the temporal modulation of interaction strengths and the rewiring of ecological topology. If $H$ are drawn from a fundamentally different interaction architecture than $B$, environmental forcing will simultaneously alter both the strength and the nature of species interactions. In that case, it would become impossible to disentangle whether observed changes in stability arise from time-dependent interaction intensities or from changes in the underlying network structure itself. By sampling $B$ and $H$ from the same ecological ensemble, we deliberately isolate the effects of environmental forcing while preserving the ecosystem's intrinsic topology, thereby allowing changes in stability to be attributed solely to the temporal modulation of existing statistical interactions.

For completeness, Appendix \eqref{Appendix} also examines the special case $H=B$. We show that, in this limit, all instantaneous community matrices commute, allowing the non-autonomous system to be solved exactly. Consequently, the periodic environmental forcing influences only the transient dynamics and has no effect on the asymptotic maximal Lyapunov exponent, irrespective of the forcing amplitude or frequency. This result provides a mathematical justification for modeling $B$ and $H$ as distinct matrices. The exactly solvable commutative limit $H=B$ eliminates the frequency dependence of the asymptotic Lyapunov exponent, suggesting that deviations from this limit are essential for capturing the nontrivial dynamical effects investigated in this study.


\subsection{Numerical Methodology}

The stability of the system is quantified by the maximal Lyapunov exponent, $\Lambda$, which measures the average exponential rate of divergence or convergence of nearby trajectories in the phase space. Negative values of $\Lambda$ indicate exponential decay of perturbations and asymptotic stability, whereas positive values correspond to exponential perturbation growth and instability. {\color{black}Because the present system is a homogeneous linear perturbation system,
$\Lambda$ is used specifically as a criterion for local asymptotic stability. A negative maximal Lyapunov exponent implies decay of all perturbation directions, whereas a positive maximal Lyapunov exponent implies the existence of an exponentially growing perturbation direction. Note that, the maximal Lyapunov exponent is not used here as a measure of chaos or as a direct measure of species persistence.} Because the considered system is driven by a time-varying matrix, analytical solutions are generally intractable outside of specific limits, necessitating rigorous numerical integration. The maximal Lyapunov exponent is computed using direct numerical integration of Eq.\ \eqref{eq:main_dynamics} combined with periodic renormalization of perturbation vectors using the standard Benettin renormalization algorithm \cite{benettin1980lyapunov1,benettin1980lyapunov2}.

Starting from a random initial condition $\mathbf{x}_0$ drawn from a standard normal distribution, the system is integrated using the adaptive Runge-Kutta-Fehlberg (RK45) algorithm implemented in the standard \texttt{SciPy} library with dynamic step sizing. {\color{black}The adaptive RK45 integration uses a relative tolerance $\mathrm{rtol}=10^{-8}$ and an absolute tolerance $\mathrm{atol}=10^{-10}$. These tolerances control the estimated local integration error during adaptive time stepping. No manually imposed maximum step size is used; the solver determines its internal step size adaptively.} Adaptive time stepping is particularly important for accurately resolving the dynamics when the forcing frequency $\omega$ becomes large and the interaction matrix varies rapidly in time. However, directly integrating exponentially growing perturbations leads rapidly to numerical overflow. To circumvent numerical overflow, we implement the standard Benettin renormalization algorithm, in which the perturbation vector is periodically rescaled to unit norm while the logarithmic growth factors are accumulated. 

The perturbation vector $\mathbf{x}(t)$ is initialized with random components and normalized to unity ($\|\mathbf{x}(0)\| = 1$). After each integration interval $\Delta t$, the perturbation vector is renormalized according to
\[
\mathbf{x}
\rightarrow
\frac{\mathbf{x}}
{\|\mathbf{x}\|}
\]
and the corresponding logarithmic growth factor is accumulated.

The system is integrated over an initial period of $t = 100$ time units, with the vector $\mathbf{x}(t)$ being renormalized to a magnitude of $1.0$ at discrete intervals of $\Delta t = 1.0$. This transient phase allows the perturbation vector to align with the dominant Lyapunov direction, thereby eliminating dependence on the initial perturbation. After this transient period, the system is integrated over a long evaluation time of $T_{max} = 7000$. At regular intervals of $\Delta t = 1.0$, the norm of the vector, $N_k = \|\mathbf{x}(t_k)\|$, is recorded, and the vector is immediately renormalized as $\mathbf{x}(t_k) \leftarrow \mathbf{x}(t_k) / N_k$ to prevent overflow. 

The maximal Lyapunov exponent is then computed as the time-averaged sum of the logarithmic growth factors:
\begin{equation}
\Lambda
=
\frac{1}{T_{\max}}
\sum_k
\ln(N_k).
\end{equation}
Because the perturbation vector is renormalized to unit norm after each integration interval, the accumulated growth factors are equivalent to the logarithm of the instantaneous norm increase between successive renormalizations. A computed value of $\Lambda < 0$ denotes a dynamically stable ecosystem that is robust to the applied periodic environmental forcing, whereas $\Lambda > 0$ indicates a dynamically unstable system.

\begin{figure*}[t]
\centering
\includegraphics[width=\textwidth]{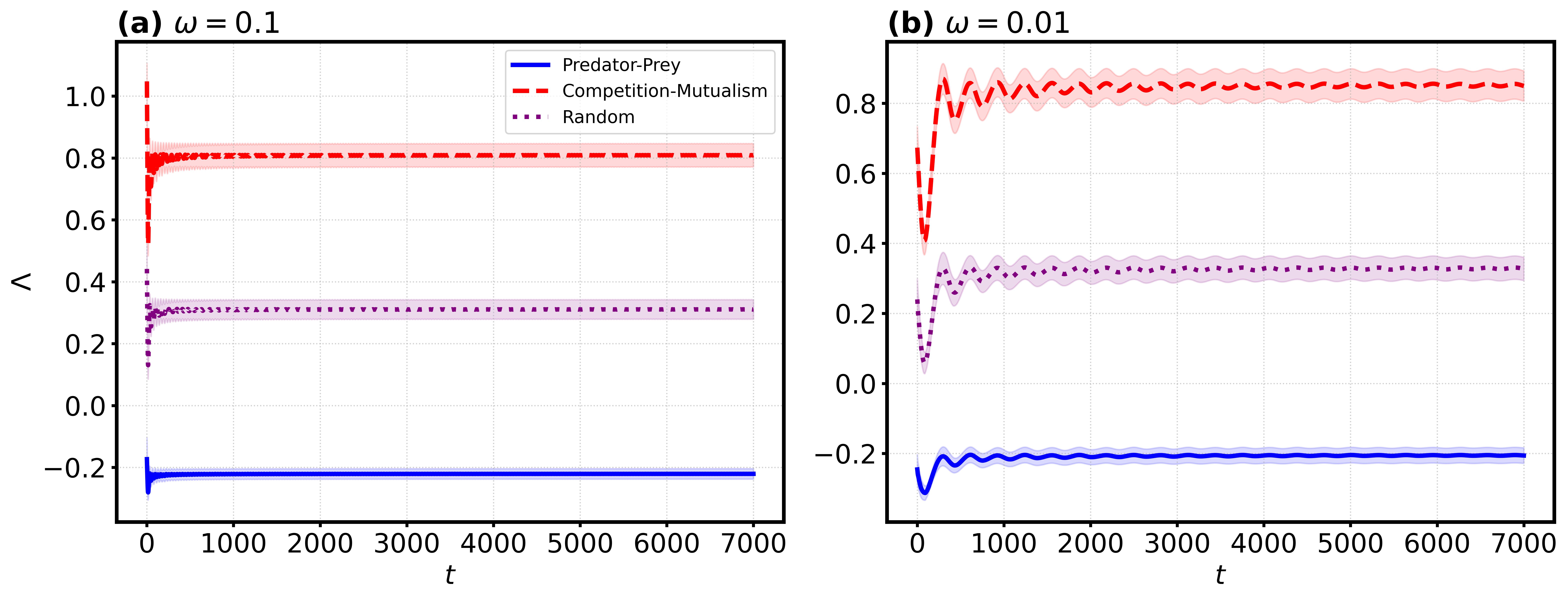}
\caption{
{\bf Convergence of the numerically computed maximal Lyapunov exponent $\Lambda$ for predator-prey (blue solid line), competition-mutualism (red dashed line), and random (purple dotted line) ecological networks obtained using the Benettin renormalization algorithm.} Panels (a) and (b) show the running estimate of the maximal Lyapunov exponent as a function of measurement time for forcing frequencies (a) $\omega=0.1$ and (b) $\omega=0.01$, respectively. Before the Lyapunov exponent is evaluated, the tangent vector is evolved for $100$ renormalization steps to align with the dominant growth direction. The running Lyapunov exponent is then computed and averaged over $30$ independent network realizations; the shaded regions indicate the standard deviation across the $30$ independent network realizations. In both cases, the ensemble-averaged running Lyapunov exponent converges to a well-defined asymptotic value, demonstrating the robustness and numerical reliability of the computation. Predator-prey communities converge to negative values of $\Lambda$, indicating asymptotic stability, whereas competition-mutualism community exhibit the largest positive Lyapunov exponents and are therefore the least stable. Random interaction networks display intermediate stability. Lower forcing frequencies produce larger transient oscillations and slower convergence because the environmental modulation acts over longer timescales. The convergence of the ensemble-averaged Lyapunov exponent by the end of the integration confirms that the chosen measurement time is sufficient for reliable estimation of the maximal Lyapunov exponent. Other parameters are fixed at $S=100$, $d=0.5$, $\sigma=0.8$, $C=0.4$, and $\varepsilon=2.0$.
}
\label{fig:Figure_7}
\end{figure*}

{\color{black}Unless otherwise stated, simulations employ a community size of $S=100$, a total evaluation time of $T_{\max}=7000$, and a renormalization interval $\Delta t=1$. The convergence of the running Lyapunov exponent with integration time is illustrated in Fig.\ \eqref{fig:Figure_7}, while the dependence on community size is examined in Fig.\ \eqref{fig:Figure_1}.} $T_{\mathrm{max}}$ should substantially exceed the forcing period $2\pi/\omega$ in order to sample many forcing cycles and obtain a converged estimate of the maximal Lyapunov exponent (See Fig.\ \eqref{fig:Figure_7}).

\section{Analytical Approximation for the Maximal Lyapunov Exponent} \label{analytical sec}

\subsection{Assumptions and Large-System Approximation}

In this section, we derive an analytical approximation for the maximal
Lyapunov exponent of the periodically forced ecological system defined
by Eq.~(\ref{eq:main_dynamics}). The derivation is based on several
simplifying assumptions that allow the application of random matrix
theory and adiabatic arguments.

Specifically, we assume:

\begin{enumerate}
    \item The community size $S$ is sufficiently large, such that
    finite-size fluctuations become negligible and standard random matrix
    results apply.

    \item {\color{black}The environmental forcing varies slowly compared with the intrinsic dynamical evolution of the system. For the parameterization used here, we consider the low-frequency regime $\omega\ll1$. For a fixed instantaneous matrix family with a nonvanishing spectral gap and bounded eigenvector variation, this limit makes the eigenvector-rotation correction asymptotically small. A more precise criterion is given later.}


    \item The spectral edge of the interaction matrices is accurately described by the large $S$ limit of random matrix theory. In particular, the leading eigenvalue is approximated by the elliptic-law boundary \cite{sommers1988spectrum,allesina2012stability}
$$
\lambda_{\max} \approx (1+\tau)\sigma\sqrt{C}. 
$$
Here,
$$
\tau = \frac{\mathbb{E}[A_{ij}A_{ji}]}{\mathrm{Var}(A_{ij})},
$$ The value of $\tau$ depends on the underlying ecological interaction architecture.

\end{enumerate}

\subsection{Structural Correlation of the Instantaneous Community Matrix}

To demonstrate that the instantaneous matrix $A(t)$ perfectly preserves the structural correlation coefficient $\tau$ of the underlying matrices, we evaluate its off-diagonal elements:
\[
A_{ij}(t) = B_{ij} + \varepsilon \cos(\omega t)H_{ij}.
\]
The correlation coefficient for this full time-dependent matrix is defined as
\[
\tau_{A} = \frac{\mathbb{E}[A_{ij}A_{ji}]}{\mathrm{Var}(A_{ij})}.
\]
Expanding the numerator, we obtain
\begin{align*}
\mathbb{E}[A_{ij}A_{ji}] &= \mathbb{E} \left[ \left(B_{ij} + \varepsilon \cos(\omega t)H_{ij}\right) \left(B_{ji} + \varepsilon \cos(\omega t)H_{ji}\right) \right].
\end{align*}
Because the baseline interaction matrix $B$ and the environmental susceptibility matrix $H$ are generated entirely independently of one  and they both have mean zero, all cross-terms (such as $\mathbb{E}[B_{ij}H_{ji}]$) are strictly zero. This reduces the expectation to
\[
\mathbb{E}[A_{ij}A_{ji}] = \mathbb{E}[B_{ij}B_{ji}] + \varepsilon^2 \cos^2(\omega t) \mathbb{E}[H_{ij}H_{ji}].
\]
By construction, $B$ and $H$ share the exact same interaction topology {\color{black}with different realizations of the connectivity pattern} and consequently possess the same structural correlation coefficient, $\tau$. We can therefore express their respective expectations in terms of their individual variances:
\begin{align*}
\mathbb{E}[B_{ij}B_{ji}] &= \tau \cdot \mathrm{Var}(B_{ij}), \\
\mathbb{E}[H_{ij}H_{ji}] &= \tau \cdot \mathrm{Var}(H_{ij}).
\end{align*}
Substituting these relations back into the expanded numerator yields
\begin{align*}
\mathbb{E}[A_{ij}A_{ji}] &= \tau \cdot \mathrm{Var}(B_{ij}) + \varepsilon^2 \cos^2(\omega t) \cdot \tau \cdot \mathrm{Var}(H_{ij}) \\
&= \tau \left[ \mathrm{Var}(B_{ij}) + \varepsilon^2 \cos^2(\omega t) \mathrm{Var}(H_{ij}) \right].
\end{align*}
Recognizing that the term inside the brackets is precisely the formulation for the variance of the instantaneous matrix elements, $\mathrm{Var}(A_{ij})$, the equation simplifies to
\[
\mathbb{E}[A_{ij}A_{ji}] = \tau \cdot \mathrm{Var}(A_{ij}).
\]
Dividing both sides by $\mathrm{Var}(A_{ij})$ confirms that the full time-dependent community matrix perfectly inherits the correlation coefficient of its constituents:
\[
\tau_A = \tau.
\]

Because the complete time-dependent community matrix $A(t)$ perfectly inherits the structural correlation of its underlying components, we can derive the correlation coefficient $\tau$ directly from the baseline interaction matrix $B$. The structural correlation coefficient is defined as
$$
\tau = \frac{\mathbb{E}[B_{ij}B_{ji}]}{\mathrm{Var}(B_{ij})}, \qquad i \neq j.
$$

For all three network architectures, the overall mean of the off-diagonal elements is zero ($\mathbb{E}[B_{ij}] = 0$), and {\color{black}the conditional variance is $\mathrm{Var}(B_{ij}) = \mathbb{E}[B_{ij}^2] = \frac{\sigma^2}{S}$. (For all three network architectures, the conditional variance of a
non-zero off-diagonal interaction is
\[
\operatorname{Var}(B_{ij}\mid B_{ij}\neq0)=\frac{\sigma^2}{S},
\]
while the unconditional variance, including the probability $(1-C)$ of
a zero interaction, is
\[
\operatorname{Var}(B_{ij})=\frac{C\sigma^2}{S}.\])} The value of $\tau$ is therefore strictly determined by the expected value of the product of reciprocal interactions, $\mathbb{E}[B_{ij}B_{ji}]$.

\paragraph{random communities.}
For completely random communities, $B_{ij}$ and $B_{ji}$ are drawn independently from a Gaussian distribution. Therefore, the expectation of their product factors perfectly:
$$
\mathbb{E}[B_{ij}B_{ji}] = \mathbb{E}[B_{ij}]\mathbb{E}[B_{ji}] = 0.
$$
This immediately yields
$$
\tau_{\mathrm{Random}} = 0.
$$

\paragraph{predator-prey communities.}
For predator-prey communities, interactions are explicitly assigned opposite signs such that $(B_{ij}, B_{ji}) = (+g_1, -g_2)$ or $(-g_1, +g_2)$, where $g_1$ and $g_2$ are independent random variables drawn from a half normal distribution, $g \sim \left| \mathcal{N}\left(0, \frac{\sigma^2}{S}\right) \right|$. 

The product of reciprocal interactions is strictly negative: $B_{ij}B_{ji} = -g_1 g_2$. Because $g_1$ and $g_2$ are independent, the expectation of their product is
$$
\mathbb{E}[B_{ij}B_{ji}] = -\mathbb{E}[g_1]\mathbb{E}[g_2].
$$
The expected value of a half normal random variable is given by $\mathbb{E}[g] = \sigma \sqrt{\frac{2}{\pi S}}$. Substituting this yields
$$
\mathbb{E}[B_{ij}B_{ji}] = - \left( \sigma \sqrt{\frac{2}{\pi S}} \right)^2 = - \frac{2}{\pi} \frac{\sigma^2}{S}.
$$

Dividing by the variance $\mathrm{Var}(B_{ij}) = \frac{\sigma^2}{S}$, we obtain the correlation coefficient:
$$
\tau_{\mathrm{PP}} = -\frac{2}{\pi}.
$$

\paragraph{competition-mutualism community.}
For these communities, the pairs that interact share the same sign, meaning $B_{ij}B_{ji} = g_1 g_2$. Following the exact same derivation for the half normal variables, the expectation is strictly positive:
$$
\mathbb{E}[B_{ij}B_{ji}] = \mathbb{E}[g_1]\mathbb{E}[g_2] = \frac{2}{\pi} \frac{\sigma^2}{S}.
$$
This yields a positive correlation coefficient:
$$
\tau_{\mathrm{CM}} = \frac{2}{\pi}.
$$

\vspace{1em}

\subsection{Adiabatic Approximation for the Maximal Lyapunov Exponent}

Under the above-mentioned assumptions, the maximal Lyapunov exponent can be
approximated analytically by averaging the instantaneous growth rate
over one forcing period.

Introducing the slow phase variable
\[
\theta=\omega t,
\]
and considering the adiabatic limit ($\omega\ll1$), the interaction matrix evolves slowly compared with the intrinsic dynamics of the ecological community. The community matrix can therefore be written as
\[
A(\theta) = -dI + B + \varepsilon\cos(\theta)H.
\]

{\color{black}More generally, the instantaneous matrix need not have time-independent
eigenvectors. Assuming that $A(\theta)$ remains diagonalizable over the
forcing cycle, we write
\[
A(\theta)=V(\theta)D(\theta)V^{-1}(\theta),
\qquad \theta=\omega t,
\]
where $D(\theta)$ contains the instantaneous eigenvalues and
$V(\theta)$ the corresponding right eigenvectors. Introducing the
instantaneous eigenbasis,
\[
x(t)=V(\theta)y(t),
\]
gives
\[
\dot y=
\left[
D(\theta)
-\omega V^{-1}(\theta)
\frac{\partial V(\theta)}{\partial\theta}
\right]y.
\]
The second term represents the eigenvector-rotation contribution. It
vanishes when the instantaneous eigenvectors are independent of time, but
is generally nonzero when $[B,H]\neq0$. It can therefore couple
instantaneous eigenmodes, so the instantaneous eigenvalues alone do not
generally determine the exact maximal Lyapunov exponent.

We order the instantaneous eigenvalues according to their real parts,
\[
\operatorname{Re}\lambda_1(\theta)
\ge
\operatorname{Re}\lambda_2(\theta)
\ge\cdots,
\]
and define the instantaneous real-part gap of an isolated dominant mode as
\[
\Delta(\theta)
=
\operatorname{Re}\lambda_1(\theta)
-
\operatorname{Re}\lambda_2(\theta)>0.
\]
A dimensionless adiabaticity parameter is then
\[
\eta(\theta)=
\frac{
\omega
\left\|
V^{-1}(\theta)
\partial_\theta V(\theta)
\right\|
}{
\Delta(\theta)
}.
\]

Note that
\[
\eta(\theta)=\omega K(\theta),
\]
where
\[
K(\theta)=
\frac{\left\|V^{-1}(\theta)\partial_\theta V(\theta)\right\|}
{\Delta(\theta)}.
\]
For a fixed instantaneous matrix family, if
\[
\sup_{\theta}K(\theta)<\infty,
\]
then
\[
\sup_{\theta}\eta(\theta)
=
\omega\sup_{\theta}K(\theta)
\longrightarrow 0
\qquad\text{as }\omega\to0.
\]
Thus, in the asymptotic low-frequency limit, $\omega\to0$
implies $\eta(\theta)\to0$, provided that the instantaneous
eigenvalue gap remains nonzero and the eigenvector variation remains
bounded throughout the forcing cycle.

The adiabatic approximation requires $\eta(\theta)\ll1$ throughout the
forcing cycle, together with an isolated dominant instantaneous mode.
Because the matrices are generally non-normal, the conditioning of
$V(\theta)$ is also relevant; the expression above should therefore be
regarded as a criterion rather than a rigorous universal error
bound.}

Under the adiabatic approximation, the matrix $A(\theta)$ changes sufficiently slowly that, over a short time interval, the system behaves as if the matrix were effectively constant. Consequently, the instantaneous growth rate of perturbations is determined by the rightmost eigenvalue of $A(\theta)$, which we denote by
\[
\lambda_{\max}(\theta).
\]
Since the matrices $B$ and $H$ are independently generated with zero mean, the variance of an off-diagonal element of $A(\theta)$ is given by
\[
\mathrm{Var}\left[A_{ij}(\theta)\right] = \mathrm{Var}(B_{ij}) + \varepsilon^2\cos^2(\theta) \mathrm{Var}(H_{ij}).
\]
Using
\[
\mathrm{Var}(B_{ij}) = \mathrm{Var}(H_{ij}) = C\frac{\sigma^2}{S},
\]
we obtain
\[
\mathrm{Var}\left[A_{ij}(\theta)\right] = C\frac{\sigma^2}{S} \left( 1+\varepsilon^2\cos^2\theta \right).
\]

Let, the effective interaction strength is given by
\[
\sigma_{eff}=\sigma\sqrt{1+\varepsilon^2\cos^2\theta}.
\]

Under the above-mentioned approximations, the rightmost edge of the eigenvalue spectrum, i.\ e.\ , the eigenvalue with the largest real part, is therefore given by
$$
R(\theta) = (1+\tau)\sigma_{\mathrm{eff}}(\theta)\sqrt{C},
$$
where $\tau$ is defined previously.

Since the community matrix contains the self-regulation term $-dI$, all eigenvalues are shifted by $-d$. Consequently, the instantaneous leading eigenvalue becomes
$$
\lambda_{\max}(\theta) = -d + (1+\tau)\sigma\sqrt{C} \sqrt{ 1+\varepsilon^2\cos^2\theta }.
$$
Within the adiabatic approximation, the matrix varies sufficiently slowly that the perturbation dynamics are governed by the instantaneous leading eigenvalue. The solution may therefore be approximated as
$$
\mathbf{x}(t) \sim \exp \left[ \int_0^t \lambda_{\max}(\omega s) \, ds \right].
$$
Taking the logarithm of the perturbation norm yields
$$
\ln \|\mathbf{x}(t)\| = \int_0^t \lambda_{\max}(\omega s) \, ds.
$$

The maximal Lyapunov exponent is then obtained from its standard definition,
$$
\Lambda = \lim_{t\rightarrow\infty} \frac{1}{t} \ln \|\mathbf{x}(t)\|,
$$
which gives
$$
\Lambda = \lim_{t\rightarrow\infty} \frac{1}{t} \int_0^t \lambda_{\max}(\omega s) \, ds.
$$
Introducing the variable
$$
\theta=\omega s,
$$
we obtain
$$
\Lambda = \lim_{T\rightarrow\infty} \frac{1}{T} \int_0^T \lambda_{\max}(\theta) \, d\theta,
$$
where the final expression represents the time average of the instantaneous leading eigenvalue over many forcing cycles.

Because the forcing enters through $\cos\theta$, which is $2\pi$-periodic, the instantaneous community matrix satisfies
\[
A(\theta+2\pi) = A(\theta),
\]
and consequently their characteristic polynomials satisfy
\[
\det\left(A(\theta+2\pi)-\lambda I\right) = \det\left(A(\theta)-\lambda I\right).
\]
Therefore, the matrices $A(\theta+2\pi)$ and $A(\theta)$ possess identical eigenvalue spectra. In particular,
\[
\lambda_{\max}(\theta+2\pi) = \lambda_{\max}(\theta).
\]

Therefore, the long-time average of the instantaneous growth rate is equal to its average over a single forcing period. Consequently,
$$
\Lambda = \lim_{T\rightarrow\infty} \frac{1}{T} \int_0^T \lambda_{\max}(\theta) \, d\theta = \frac{1}{2\pi} \int_0^{2\pi} \lambda_{\max}(\theta) \, d\theta.
$$
Substituting the expression for the instantaneous leading eigenvalue,
$$
\lambda_{\max}(\theta) = -d + (1+\tau)\sigma\sqrt{C} \sqrt{1+\varepsilon^2\cos^2\theta},
$$
gives

\begin{equation}
\Lambda = -d + (1+\tau)\sigma\sqrt{C} M(\varepsilon),
\label{eq:lambda_analytical}
\end{equation}
where
\begin{equation}
M(\varepsilon) = \frac{1}{2\pi} \int_0^{2\pi} \sqrt{ 1+\varepsilon^2\cos^2\theta } \, d\theta.
\label{eq:M_epsilon}
\end{equation}

Equation above provides an analytical approximation for the maximal Lyapunov exponent in the adiabatic limit ($\omega\ll1$). {\color{black}Equation \eqref{eq:lambda_analytical} should therefore be interpreted as the leading adiabatic approximation rather than an exact identity for a general non-normal, non-commuting matrix family.} The function $M(\varepsilon)$ quantifies the average amplification induced by periodic environmental forcing and depends solely on the forcing amplitude $\varepsilon$. Since the integrand of $\frac{dM(\varepsilon)}{d\varepsilon}$ is non-negative and vanishes only on a set of measure zero, it follows that $M(\varepsilon)$ is a strictly increasing function of $\varepsilon$. Moreover, given that $M(0)=1$, it holds that $M(\varepsilon) \ge 1$ for all $\varepsilon \ge 0$, with equality attained only at $\varepsilon=0$. Consequently, {\it within the adiabatic approximation, increasing the forcing amplitude monotonically increases the predicted maximal Lyapunov exponent.}

Furthermore, Eq.\ \eqref{eq:lambda_analytical} provides an explicit mathematical condition for local perturbation stability. The maximal Lyapunov exponent is negative (indicating a stable ecosystem) if and only if the intrinsic self-regulation overcomes the environmentally amplified interaction network, specifically requiring$$d > (1+\tau)\sigma\sqrt{C}M(\varepsilon).$$Because the destabilizing term on the right-hand side scales directly with $(1+\tau)$, the specific interaction architecture fundamentally dictates how easily this stability condition is satisfied.  {\color{black} Thus, within the adiabatic large-system approximation, $d_c=(1+\tau)\sigma\sqrt{C}M(\varepsilon)$ is the critical self-regulation rate separating $\Lambda<0$ from $\Lambda>0$. The threshold depends on interaction topology through $\tau$, interaction strength through $\sigma$, connectance through $C$, and forcing amplitude
through $M(\epsilon)$. This is a local perturbation-stability threshold and should not be interpreted as a threshold for species persistence or population abundance.}

\subsection{Topology-Dependent Stability Predictions}

The analytical approximation obtained above immediately reveals how interaction topology influences ecological stability through the correlation coefficient $\tau$. For the three network architectures considered in this study,
\[
\tau_{\mathrm{PP}} = -\frac{2}{\pi}, \qquad \tau_{\mathrm{Random}} = 0, \qquad \tau_{\mathrm{CM}} = \frac{2}{\pi},
\]
corresponding to predator-prey, random, and competition-mutualism community, respectively.

Substituting these values into Eq.~(\ref{eq:lambda_analytical}) yields
\[
\Lambda_{\mathrm{PP}} = -d + \left(1 - \frac{2}{\pi}\right)\sigma\sqrt{C} M(\varepsilon),
\]
for predator-prey communities,
\[
\Lambda_{\mathrm{Random}} = -d + \sigma\sqrt{C} M(\varepsilon),
\]
for random communities, and
\[
\Lambda_{\mathrm{CM}} = -d + \left(1 + \frac{2}{\pi}\right)\sigma\sqrt{C} M(\varepsilon),
\]
for competition-mutualism communities.

These expressions immediately predict a strict ordering of stability,
\[
\Lambda_{\mathrm{PP}} < \Lambda_{\mathrm{Random}} < \Lambda_{\mathrm{CM}},
\]
for any fixed set of parameters $(d,\sigma,C,\varepsilon)$. Since negative values of $\Lambda$ correspond to stable dynamics, predator-prey communities are predicted to be the most stable ecological architecture, while competition-mutualism community are expected to be the least stable. random communities occupy an intermediate position between these two extremes. It is important to emphasize that Eq.\ \eqref{eq:lambda_analytical} is derived under the adiabatic assumption $(\omega \ll 1)$, which allows the community matrix to be treated as effectively constant over short time intervals. At larger forcing frequencies, dynamical averaging effects become important. Consequently, deviations between the analytical prediction and numerical simulations are expected outside the adiabatic regime.

The observed hierarchy emerges directly from the interaction structure encoded in $\tau$. In predator-prey systems, opposite-sign interactions generate negative correlations between reciprocal interaction pairs, suppressing the spectral edge of the interaction matrix and thereby enhancing stability. In contrast, competition-mutualism community contain positively correlated reciprocal interactions that enlarge the spectral edge and promote instability. Random communities lack any systematic reciprocal correlation and therefore exhibit intermediate behavior.

\begin{figure*}[t]
\centering
\includegraphics[width=\textwidth]{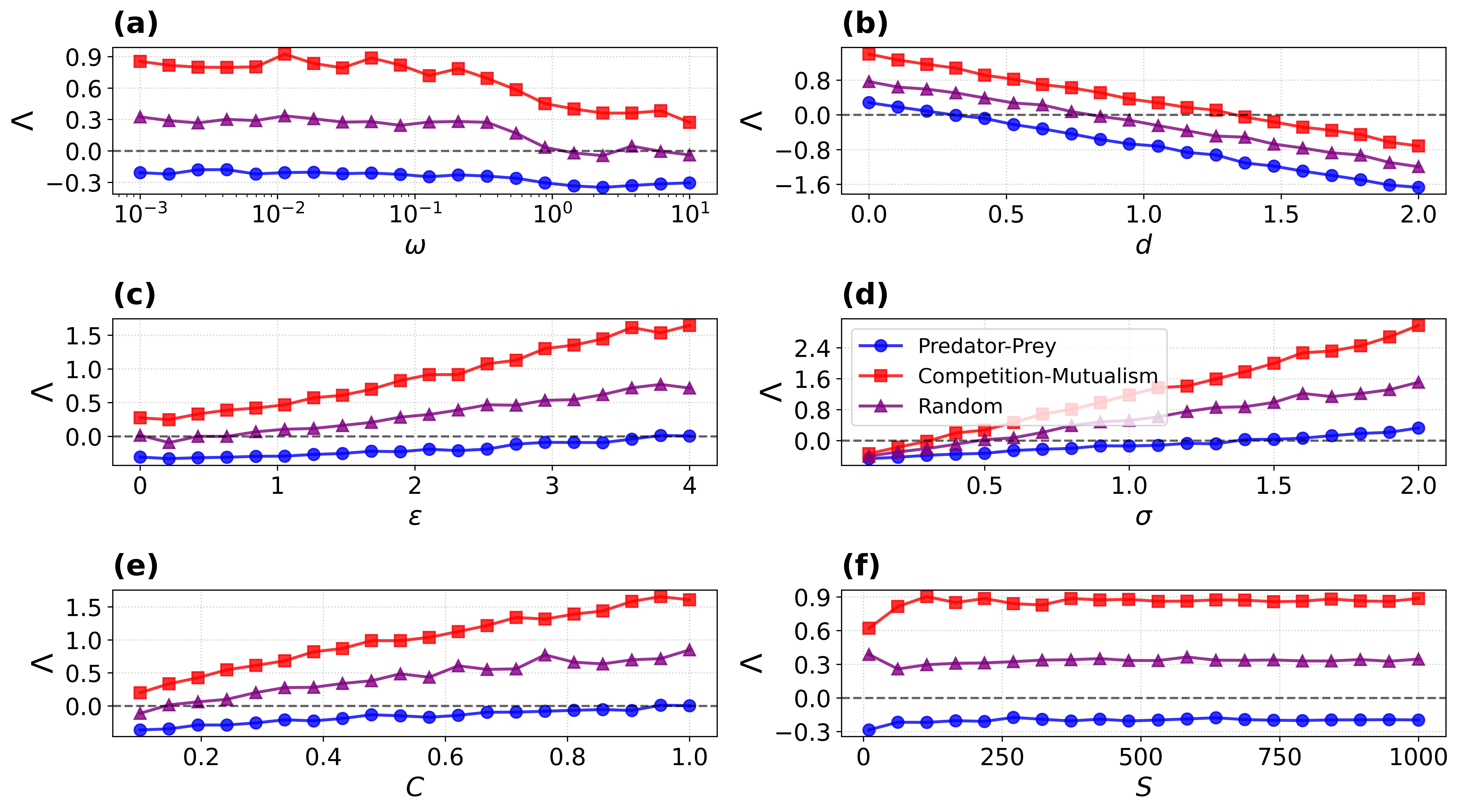}
\caption{
{\bf Dependence of the maximal Lyapunov exponent $\Lambda$ on ecological and environmental parameters for predator-prey (blue circles), competition-mutualism (red squares), and random (purple triangles) interaction networks.} (a) Increasing forcing frequency $\omega$ generally reduces instability in random and competition-mutualism communities, while predator-prey systems remain predominantly stable across the entire frequency range, particularly for these chosen parameter values. (b) Increasing self-regulation $d$ shifts $\Lambda$ toward negative values for all interaction architectures, demonstrating the stabilizing role of species-level damping. (c) Stronger environmental forcing $\varepsilon$ increases $\Lambda$ and promotes instability, particularly in competition-mutualism community, whereas predator-prey communities remain comparatively resilient for these chosen parameter values. (d) Larger interaction strengths $\sigma$ increase instability in all interaction architectures, with competition-mutualism communities exhibiting the strongest sensitivity and predator-prey communities the weakest. (e) Increasing network connectivity $C$ generally raises the maximal Lyapunov exponent, indicating that densely connected ecological communities are more susceptible to instability under periodic forcing. (f) The Lyapunov exponent exhibits only weak dependence on system size $S$. The horizontal dashed line at $\Lambda = 0$ indicates the critical transition boundary separating stable ($\Lambda < 0$) from unstable ($\Lambda > 0$) dynamics. Unless otherwise varied on the x-axis, the system parameters are strictly fixed at $S = 100$, $d = 0.5$, $\sigma = 0.8$, $C = 0.4$, $\epsilon = 2.0$, and $\omega = 0.01$. }
\label{fig:Figure_1}
\end{figure*}

\section{Numerical Results}

Furthermore, Eq.\ (\ref{eq:lambda_analytical}) predicts that {\it increasing the self-regulation strength $d$ stabilizes the ecosystem, whereas increasing the interaction strength $\sigma$, connectance $C$, or forcing amplitude $\varepsilon$ increases the maximal Lyapunov exponent and promotes instability in the adiabatic limit ($\omega\ll1$).} These theoretical predictions will be tested against direct numerical simulations in the following sections.

\begin{figure*}[t]
\centering
\includegraphics[width=\textwidth]{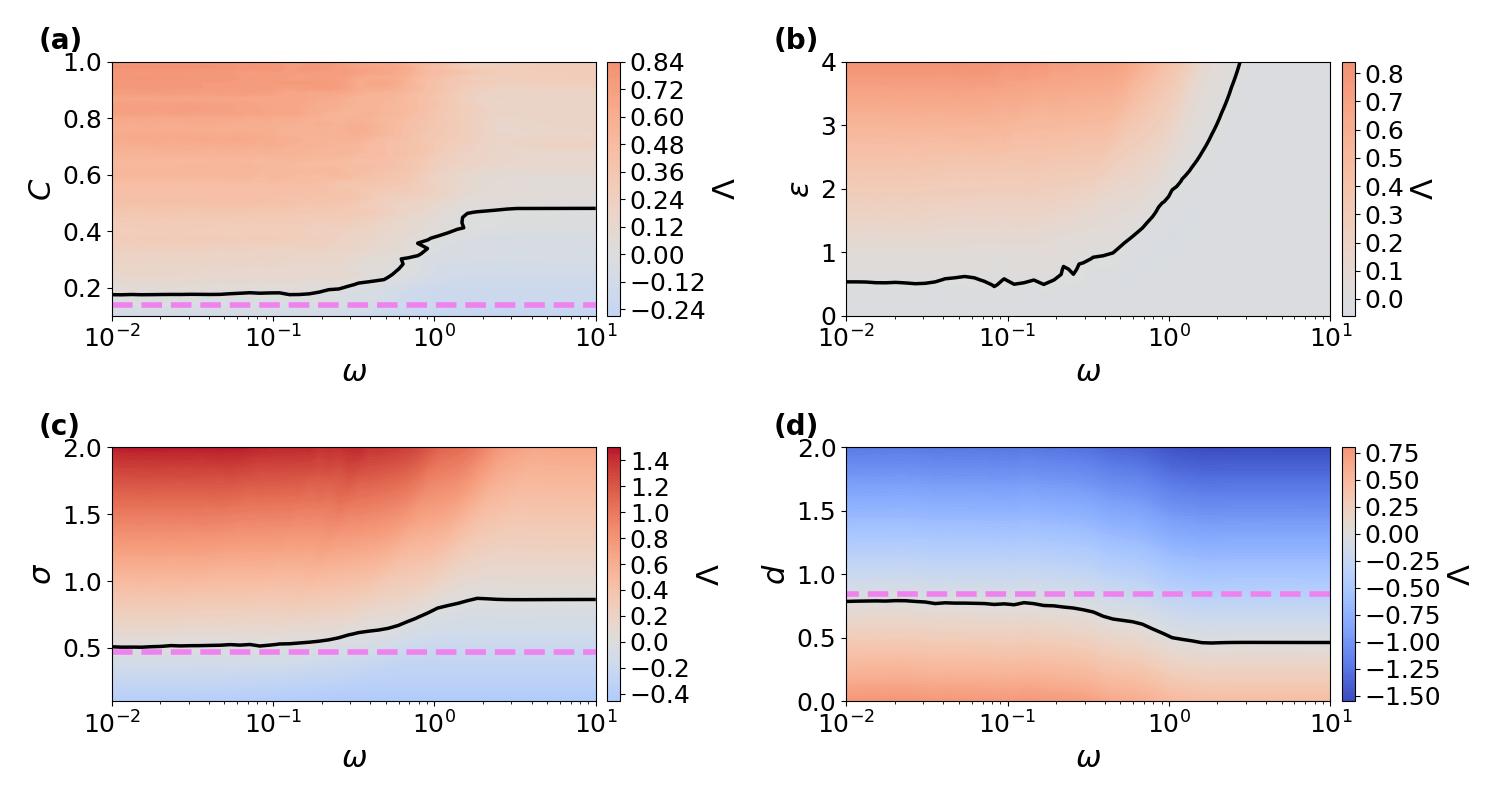}
\caption{
{\bf Stability phase diagrams for random ecological communities under periodic environmental forcing.} The color scale represents the numerically computed maximal Lyapunov exponent $\Lambda$, with $\Lambda>0$ indicating unstable dynamics, and $\Lambda<0$ indicating stable dynamics. The solid black curve denotes the numerical stability boundary defined by $\Lambda=0$, while the dashed violet curve indicates the corresponding analytical prediction obtained from Eq.\ \eqref{eq:lambda_analytical}. (a) Stability landscape in the $(C,\omega)$ plane, showing how network connectance and forcing frequency jointly determine the onset of instability. Notably increasing  (b) Stability landscape in the $(\varepsilon,\omega)$ plane. The strong upward bending of the stability boundary demonstrates that high $\omega$ substantially suppresses the destabilizing effect of large environmental fluctuations. (c) Increasing interaction strength $\sigma$ promotes instability, whereas increasing forcing frequency $\omega$ shifts the critical interaction strength required for destabilization to larger values. (d) Increasing self-regulation $d$ stabilizes the system, and rapid environmental forcing $\omega$ further enlarges the stable region of parameter space. Across all parameter planes, increasing the forcing frequency generally decreases the maximal Lyapunov exponent and expands the region of stability. The results demonstrate that the temporal scale of environmental variability plays a crucial role in determining ecosystem resilience under periodic forcing. Unless otherwise specified, all non-varying parameters are held constant at the fixed values: $\varepsilon=2.0$, $C = 0.4$, $\sigma= 0.8$, $d=0.5$, and $S=100$.
}
\label{fig:Figure_2}
\end{figure*}

\subsection{Dependence of Stability on interaction topology}

\begin{figure*}[t]
\centering
\includegraphics[width=\textwidth]{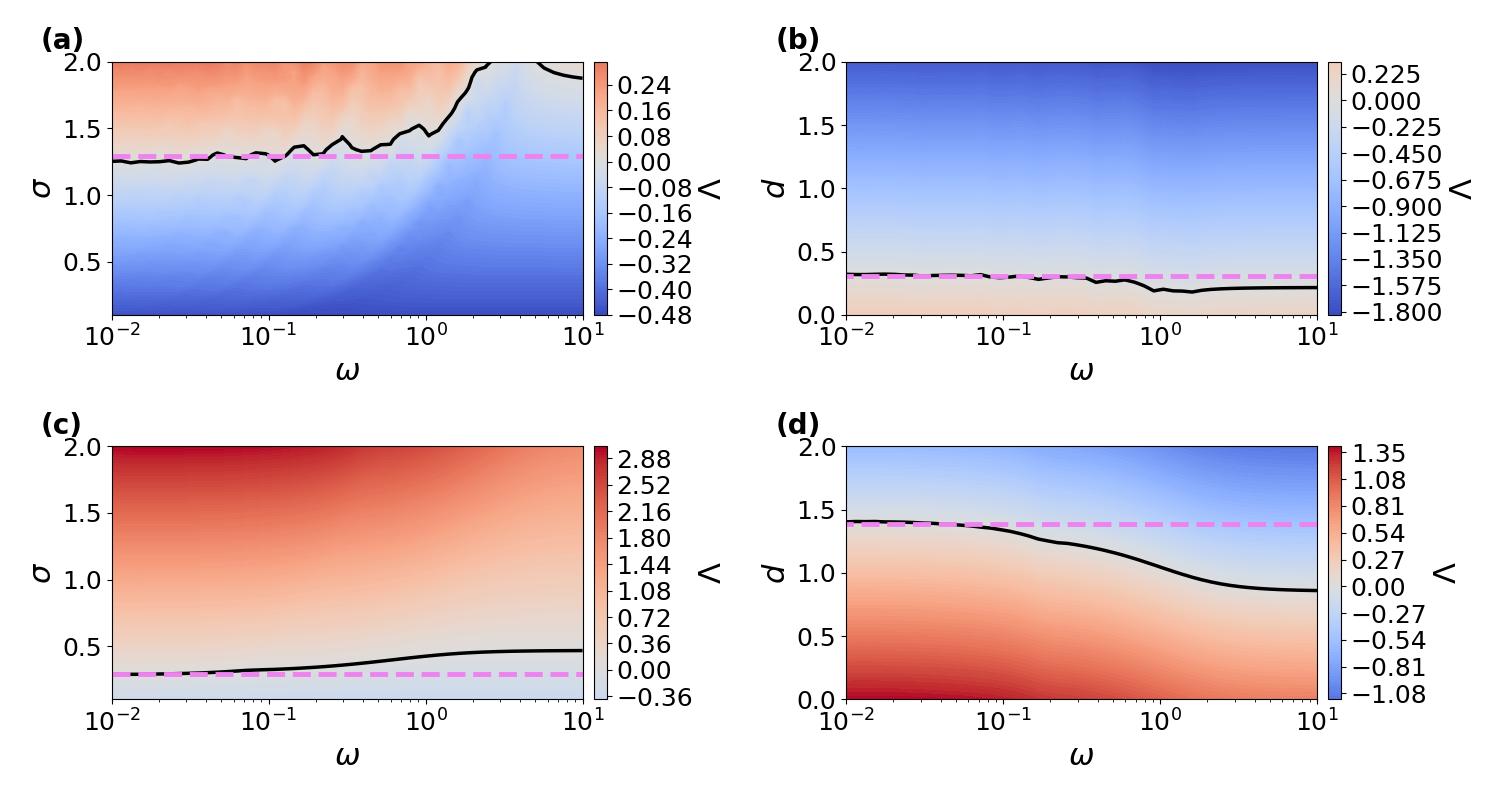}
\caption{
{\bf Stability phase diagrams for predator-prey and competition-mutualism ecological communities under periodic environmental forcing.}
The maximal Lyapunov exponent $\Lambda$ is shown as a function of forcing frequency $\omega$ and interaction strength $\sigma$ (left column) or self-regulation rate $d$ (right column). The top row corresponds to predator-prey communities [(a),(b)], while the bottom row corresponds to competition-mutualism communities [(c),(d)]. Color indicates the value of the maximal Lyapunov exponent, with $\Lambda<0$ regions corresponding to stable dynamics and $\Lambda>0$ regions corresponding to unstable dynamics. The solid black curves $\Lambda=0$ denote the numerically determined stability boundaries. The dashed curves show the analytical prediction obtained from Eq.~(\ref{eq:lambda_analytical}), which is derived under the adiabatic approximation ($\omega\ll1$). Excellent agreement is observed at low forcing frequencies, while systematic deviations emerge as $\omega$ increases and the assumptions of the analytical theory become less accurate. Panels (a) and (c) demonstrate that increasing interaction strength destabilizes both network architectures, although competition-mutualism communities remain substantially more fragile than predator-prey communities across the entire parameter range. Panels (b) and (d) show that increasing self-regulation stabilizes the community by shifting the system toward negative Lyapunov exponents. In both architectures, increasing forcing frequency enlarges the stable region, indicating a frequency-induced stabilization effect whereby rapidly varying environmental fluctuations become less effective at generating long-term perturbation growth. Unless otherwise stated, parameters are fixed at $S=100$, $C=0.4$, $\varepsilon=2.0$, $\sigma=0.8$, and $d=0.5$, with only the parameters shown on the axes varied.
}
\label{fig:topological_comparison}
\end{figure*}

A central question in theoretical ecology concerns how the architecture of species interactions influences the stability of ecological communities. To address this question, we compute the maximal Lyapunov exponent, $\Lambda$, for predator-prey, competition-mutualism, and random interaction networks under periodic environmental forcing. Figure~\ref{fig:Figure_1} summarizes the dependence of $\Lambda$ on the principal ecological and environmental parameters. In each panel, one parameter is varied while all others are fixed at their baseline values: $S=100$, $d=0.5$, $\sigma=0.8$, $C=0.4$, $\varepsilon=2.0$, and $\omega=10^{-2}$. 

Since $\Lambda<0$ corresponds to asymptotically stable dynamics and $\Lambda>0$ indicates instability, the results reveal a clear hierarchy of {\color{black}local perturbation} stability among the three network architectures. Across nearly the entire parameter range explored, predator-prey communities exhibit the smallest Lyapunov exponents and are therefore the most stable. In contrast, competition-mutualism communities consistently display the largest values of $\Lambda$, indicating the greatest susceptibility to instability. random communities occupy an intermediate position between these two extremes.

This ordering can be understood from the underlying interaction structure. Predator-prey interactions contain an inherent balance of positive and negative feedback that tends to suppress the amplification of perturbations. competition-mutualism community, by contrast, contain interaction motifs that can reinforce population fluctuations and promote instability. random communities lack a systematic interaction pattern and therefore exhibit intermediate behavior.

The influence of individual model parameters is shown in Fig.~\ref{fig:Figure_1}(a-f). Increasing the self-regulation parameter $d$ systematically decreases $\Lambda$ for all network architectures [Fig.~\ref{fig:Figure_1}(b)], confirming the stabilizing role of intrinsic damping. Conversely, increasing the forcing amplitude $\varepsilon$ [Fig.~\ref{fig:Figure_1}(c)], interaction strength $\sigma$ [Fig.~\ref{fig:Figure_1}(d)], or connectance $C$ [Fig.~\ref{fig:Figure_1}(e)] generally increases $\Lambda$ and promotes instability.

The destabilizing effect of $\varepsilon$ arises because stronger environmental forcing produces larger temporal fluctuations in the interaction matrix, thereby amplifying deviations from equilibrium trajectories. Similarly, increasing $\sigma$ strengthens species interactions and enhances the growth of perturbations. This observation is consistent with classical random-matrix arguments \cite{may1972will}, where larger interaction variances broaden the eigenvalue spectrum and increase the likelihood of instability. Increasing connectance $C$ also destabilizes the community by increasing the number of pathways through which perturbations can propagate across the network. Consequently, disturbances affecting individual species become more readily transmitted throughout the community. Figure~\ref{fig:Figure_1}(f) demonstrates that the maximal Lyapunov exponent depends only weakly on the total number of species once the system size becomes sufficiently large. Beyond a certain value of $S$, the stability measures approach nearly constant values, indicating convergence toward a large-system regime in which finite-size effects are relatively small.

A particularly notable result is shown in Fig.~\ref{fig:Figure_1}(a). As the forcing frequency $\omega$ increases, the maximal Lyapunov exponent decreases for all three network architectures. This frequency-dependent suppression of instability suggests that rapidly varying environmental fluctuations are less effective at destabilizing ecological communities than slowly varying perturbations. The origin and consequences of this stabilization mechanism are explored in greater detail in the following subsection through a systematic analysis of the stability phase diagrams.

\begin{figure*}[t]
    \centering
    \includegraphics[width=\textwidth]{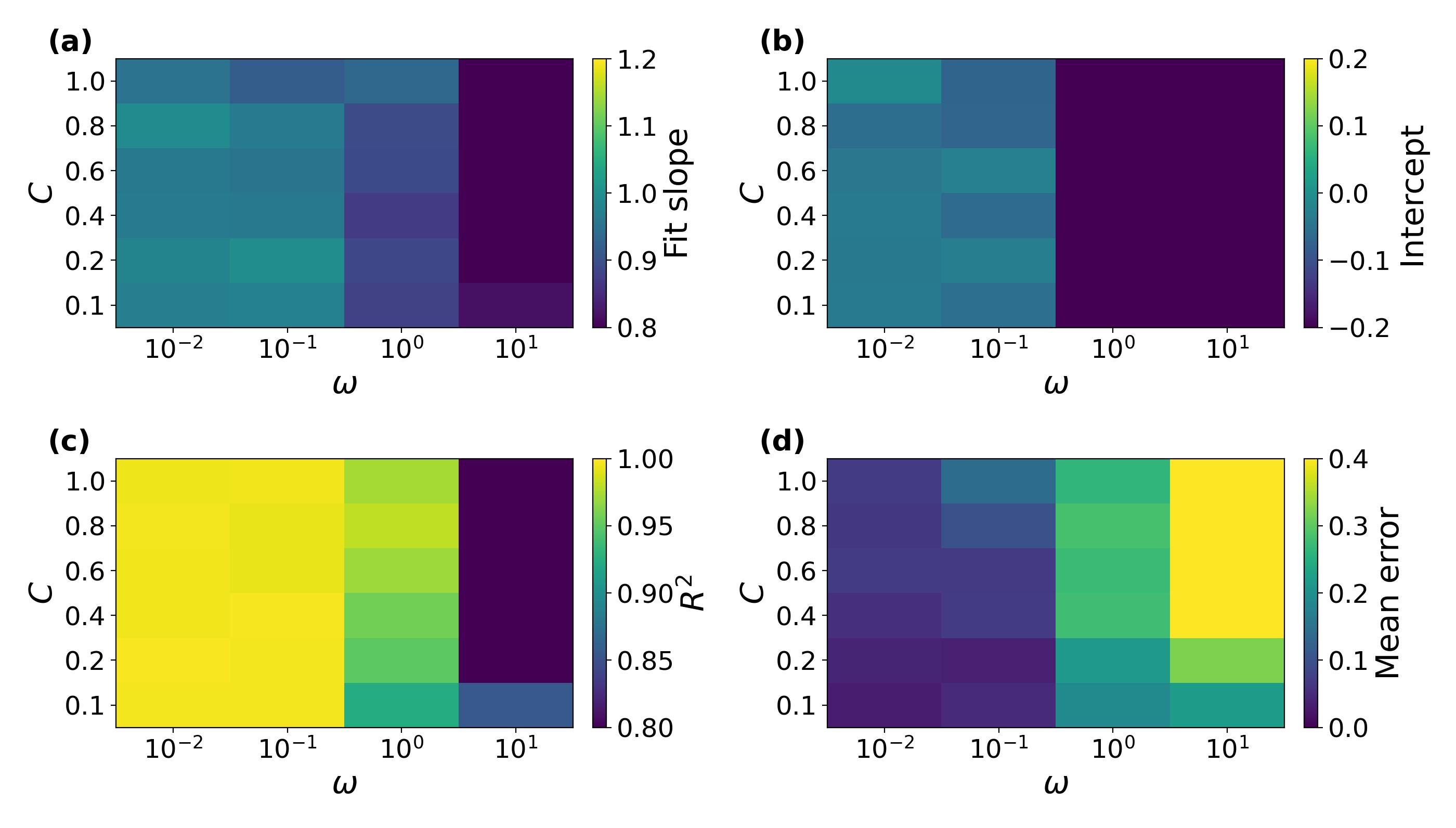}
    \caption{\textbf{Statistical validation of the adiabatic theoretical approximation for random communities.} The heatmaps display the results of a linear regression ($\Lambda_{\text{num}} = m\Lambda_{\text{th}} + c$) comparing the numerically calculated maximal Lyapunov exponent $\Lambda_{\text{num}}$ to the analytical prediction $\Lambda_{\text{th}}$ across the $(C, \omega)$ parameter plane. The panels show (a) the fit slope $m$, (b) the y-intercept $c$, (c) the coefficient of determination $R^2$, and (d) the mean absolute error $|\Lambda_{\text{num}} - \Lambda_{\text{th}}|$. Each pixel summarizes 216 parameter combinations obtained by varying $\sigma \in [0.3, 1.5]$, $d \in [0.0, 1.5]$, and $\varepsilon \in [0.0, 4.0]$. The theoretical approximation shows near-perfect agreement (slope $\approx 1$, $R^2 \approx 1$, error $\approx 0$) in the adiabatic, low-frequency limit ($\omega \le 10^{-1}$), but predictably deteriorates as high-frequency forcing invalidates the quasi-static assumption.}
    \label{fig:random_validation}
\end{figure*}

\subsection{Stability Phase Diagrams Under Periodic Forcing}

\begin{figure*}[t]
    \centering
    \includegraphics[width=\textwidth]{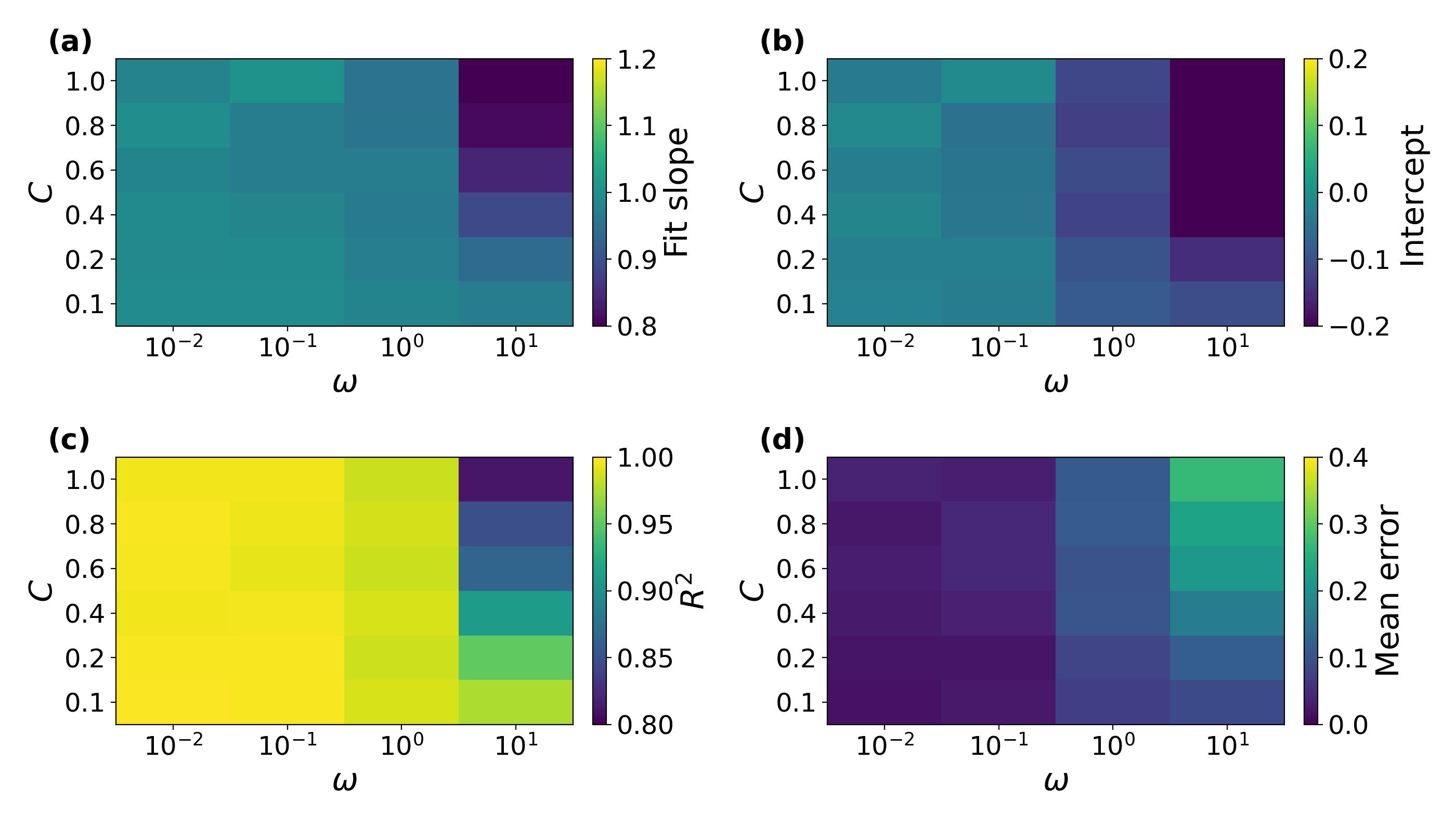}
    \caption{{\bf Statistical validation of the adiabatic theoretical approximation for predator-prey communities.} The heatmaps display the results of a linear regression ($\Lambda_{\text{num}} = m\Lambda_{\text{th}} + c$) comparing the numerically exact maximal Lyapunov exponent to the analytical prediction ($\tau_{\mathrm{PP}} = -2/\pi$) across the $(C, \omega)$ parameter plane. The panels represent (a) the fit slope $m$, (b) the y-intercept $c$, (c) the coefficient of determination $R^2$, and (d) the mean absolute error $|\Lambda_{\text{num}} - \Lambda_{\text{th}}|$. Each pixel summarizes $216$ parameter combinations obtained by sweeping $\sigma \in [0.3, 1.5]$, $d \in [0.0, 1.5]$, and $\varepsilon \in [0.0, 4.0]$ while holding the underlying matrix topology fixed for each $\sigma$. The theoretical approximation is highly accurate (slope $\approx 1$, $R^2 \approx 1$, error $\approx 0$) in the adiabatic limit ($\omega \le 10^{-1}$). The decreasing regression slope, together with the increasing mean absolute error and the reduction in $R^2$, indicates that the adiabatic approximation progressively loses quantitative accuracy as the forcing frequency increases. This behavior is consistent with the emergence of non-adiabatic dynamical effects that are not captured by the present analytical framework.}
    \label{fig:predatorprey_validation}
\end{figure*}

\begin{figure*}[t]
    \centering
    \includegraphics[width=\textwidth]{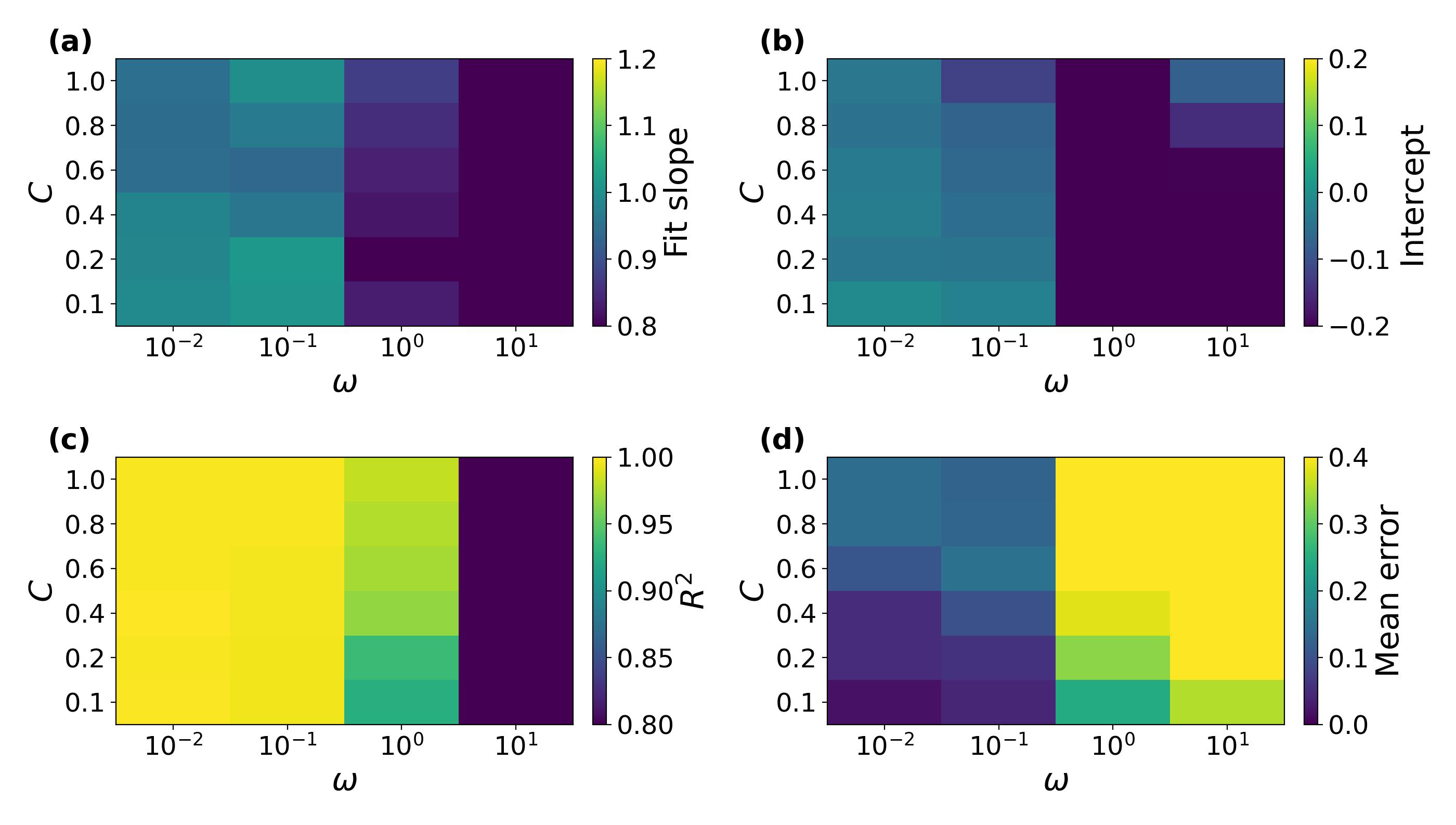}
    \caption{\textbf{Statistical validation of the adiabatic theoretical approximation for competition-mutualism community.} The heatmaps display the results of a linear regression ($\Lambda_{\text{num}} = m\Lambda_{\text{th}} + c$) comparing the numerically exact maximal Lyapunov exponent to the analytical prediction ($\tau_{\mathrm{CM}} = 2/\pi$) across the $(C, \omega)$ parameter plane. The panels display (a) the fit slope $m$, (b) the y-intercept $c$, (c) the coefficient of determination $R^2$, and (d) the mean absolute error $|\Lambda_{\text{num}} - \Lambda_{\text{th}}|$. Each pixel summarizes $216$ parameter combinations obtained by varying $\sigma \in [0.3, 1.5]$, $d \in [0.0, 1.5]$, and $\varepsilon \in [0.0, 4.0]$ while holding the underlying matrix topology fixed for each $\sigma$. Similar to the previous architectures, the theoretical approximation is highly accurate (slope $\approx 1$, $R^2 \approx 1$, error $\approx 0$) in the adiabatic limit ($\omega \le 10^{-1}$). The regression slope decreases from values close to unity, the coefficient of determination declines, and the mean absolute error increases as the forcing frequency increases. Together, these diagnostics indicate that the quantitative agreement between the adiabatic analytical approximation and the numerical simulations progressively deteriorates as the forcing frequency increases.}
    \label{fig:compmut_validation}
\end{figure*}

To examine how periodic environmental forcing modifies ecosystem stability, we construct two-dimensional stability phase diagrams for the random communities in Fig.\ \eqref{fig:Figure_2}. To ensure smooth and reproducible parameter sweeps, all simulations are performed using a fixed underlying realization of the interaction matrices. Specifically, a master set of Gaussian random variables and uniform probability matrices is generated once with a fixed random seed and subsequently reused throughout the parameter exploration. As a result, variations in the maximal Lyapunov exponent arise solely from changes in the control parameters, thereby eliminating fluctuations due to independent network sampling. For variations across the forcing frequency $\omega$, forcing amplitude $\epsilon$, and intrinsic self-regulation rate $d$, the underlying interaction matrices remain strictly unchanged. When varying the connectance $C$, links are progressively activated from the same underlying adjacency template (i.e., by comparing the uniform probability matrix against the threshold $C$), thereby preserving the existing topological backbone. Likewise, varying the interaction strength $\sigma$ uniformly rescales the non-zero matrix elements while maintaining the exact interaction topology and sign structure. This procedure rigorously isolates the effect of each parameter, yielding the smooth stability boundaries observed across the phase diagrams. {\color{black}These phase diagrams characterize the parameter dependence of the specified network realization and should not be interpreted as ensemble confidence regions or statistical uncertainty bounds.}

Figure~\ref{fig:Figure_2} shows the maximal Lyapunov exponent, $\Lambda$, across four parameter planes involving the forcing frequency $\omega$. The color scale represents the numerical Lyapunov exponent obtained from direct integration of the periodically forced community matrix model \eqref{eq:main_dynamics}, with $\Lambda>0$ indicating instability and $\Lambda<0$ indicating stability. The solid black curves denote the numerical stability boundaries defined by $\Lambda=0$, while the dashed curves correspond to the analytical prediction derived in Sec.\ \eqref{analytical sec}.

Figure \eqref{fig:Figure_2} reveals that the stability boundary exhibits a pronounced dependence on the forcing frequency. In the $(C,\omega)$ plane [Fig.~\ref{fig:Figure_2}(a)], increasing connectance destabilizes the community, shifting the system toward positive Lyapunov exponents. However, the critical connectance required for instability increases substantially as the forcing frequency becomes larger. A similar trend is observed in the $(\sigma,\omega)$ plane [Fig.~\ref{fig:Figure_2}(c)], where stronger interactions are required to destabilize rapidly forced systems. In contrast, the $(d,\omega)$ plane [Fig.~\ref{fig:Figure_2}(d)] demonstrates that increasing self-regulation stabilizes the community, with the critical damping threshold decreasing as the forcing frequency increases. Together, these results indicate that rapid environmental oscillations suppress the growth of perturbations and enlarge the stable region of parameter space. All these results are consistent with our findings in Fig.\ \eqref{fig:Figure_1}.

The $(\varepsilon,\omega)$ plane in subfigure \eqref{fig:Figure_2}(b) indicates that for sufficiently small $\omega$, increasing $\varepsilon$ promotes instability. Surprisingly, however, the destabilizing influence of large-amplitude forcing is progressively weakened as $\omega$ increases. Consequently, the stability boundary bends strongly toward larger values of $\varepsilon$, implying that rapidly oscillating environments can tolerate substantially larger fluctuations before destabilization occurs.

A noticable feature of the phase diagrams is the absence of the theoretical boundary (dashed line) in this subfigure (b), where the forcing amplitude $\varepsilon$ is varied against frequency $\omega$. This absence is a consequence of the baseline parameter regime ($d=0.5, \sigma=0.8, C=0.4$). Since the forcing frequency $\omega$ does not appear explicitly in the expression \eqref{eq:lambda_analytical}, the theoretical stability threshold
$\Lambda_{th}=0$ predicts a critical forcing amplitude that is independent of frequency. Also, $\varepsilon$ appears as a function of $M(\varepsilon)$. At these chosen fixed parameter values, our theoretical prediction suggests that the ecosystem with $\varepsilon = 0$ is already unstable with $\Lambda_{th} \approx 0.006 > 0$. Because increasing the forcing amplitude strictly amplifies the time-averaged fluctuation function $M(\epsilon)$, the large-$S$ adiabatic approximation predicts that the system remains unstable throughout the plotted parameter range of subfigure (b). 
Consequently, the analytical theory cannot reproduce the pronounced bending of the numerical stability boundary observed in the $(\varepsilon,\omega)$ plane. Furthermore, the discrepancy between the analytical prediction and the numerical stability boundary in Fig.\ \eqref{fig:Figure_2}(b) may partially be influenced by finite-size effects. The analytical approximation is derived in the asymptotic large-system limit ($S \to \infty$), whereas the numerical simulations are performed for a finite community ($S=100$). Since the parameter values considered lie very close to the theoretical stability threshold, small finite-size fluctuations in the spectral edge of the realized interaction matrix may produce noticeable shifts in the numerical stability boundary. 
The missing dashed contour, therefore, reflects a genuine limitation of the approximation, indicating that frequency-dependent stabilization arises from dynamical effects beyond the scope of the perceived analytical arguments. However, the numerical integration (black solid line) reveals a substantial region of stability ($\Lambda<0$) emerging exclusively at high frequencies ($\omega > 1$). This phenomenon highlights the most profound consequence of the environment: rapid, periodic environmental fluctuations do not merely shift existing stability boundaries; they can dynamically stabilize interaction topologies that are fundamentally unstable in static or slowly changing environments.

The numerical results suggest that this stabilization originates from a
dynamical averaging mechanism. Consider the governing equation \eqref{eq:main_dynamics}

\[
\frac{d\mathbf{x}}{dt}
=
\left(
-dI+B+\varepsilon\cos(\omega t)H
\right)\mathbf{x}.
\]

For sufficiently high forcing frequencies, the environmental forcing varies much more rapidly than the ecosystem can respond to. Consequently, the effects of successive positive and negative fluctuations tend to cancel each other over time. Averaging theory then implies that the net contribution of the periodic term becomes progressively smaller.




Indeed, for any finite time interval $[t_1, t_2]$, we have
\begin{equation}
\int_{t_1}^{t_2} \varepsilon \cos(\omega t) \, dt = \frac{\varepsilon}{\omega} \left[ \sin(\omega t_2) - \sin(\omega t_1) \right],
\end{equation}
which vanishes in the limit $\omega \to \infty$ since the sine function is bounded. {\color{black}This behavior is consistent with the Riemann--Lebesgue lemma,
which states that, for an integrable function $f(t)$,
\[
\int_a^b f(t)e^{i\omega t}\,dt \to 0
\qquad \text{as } \omega\to\infty.
\]} In the high-frequency limit, the positive and negative phases of the environmental forcing cancel over time, and the system behaves increasingly like the following system

\[
\frac{d\mathbf{x}}{dt}
\approx
(-dI+B)\mathbf{x}.
\]

As a consequence, larger forcing amplitudes are required to destabilize
the community as $\omega$ increases, producing the strong upward bending
of the numerical stability boundary observed in
Fig.~\ref{fig:Figure_2}(b).

{\color{black}

To characterize the high-frequency limit more rigorously, we write
\[
A(t)=A_0+\epsilon\cos(\omega t)H,
\qquad
A_0=-dI+B,
\]
and consider the evolution over one forcing period
\[
T=\frac{2\pi}{\omega}.
\]
The one-period propagator can be written in Floquet form \cite{klausmeier2008floquet} as
\[
U(T)=\exp[\Omega(T)],
\]
where $\Omega(T)$ is given by the Magnus expansion \cite{wiki_magnus,blanes2009magnus}

\[
\Omega=\Omega_1+\Omega_2+\Omega_3+\cdots.
\]

The first term is
\[
\Omega_1
=
\int_0^T A(t)\,dt
=
A_0T,
\]
because
\[
\int_0^T\cos(\omega t)\,dt=0.
\]

Although the first-order average is therefore $A_0$, the matrices at
different times need not commute:
\[
[A(t_1),A(t_2)]
=
\epsilon
[\cos(\omega t_2)-\cos(\omega t_1)]
[A_0,H],
\]
where $[A(t_1), A(t_2)]$ denotes the matrix commutator, defined by $[A(t_1), A(t_2)] = A(t_1)A(t_2) - A(t_2)A(t_1)$. Thus, the vanishing average of the scalar forcing term alone does not establish the high-frequency limit.

For the single-harmonic cosine forcing considered here, the second Magnus
term vanishes over one complete period, $$\Omega_2=\frac{1}{2}
\int_0^T dt_1
\int_0^{t_1}dt_2\,
[A(t_1),A(t_2)]=0.$$

The first non-vanishing correction therefore arises from the third
Magnus term,
\begin{align*}
\Omega_3
&=
\frac{1}{6}
\int_0^T dt_1
\int_0^{t_1}dt_2
\int_0^{t_2}dt_3
\Big(
[A(t_1),[A(t_2),A(t_3)]]
\nonumber\\
&\qquad\qquad\qquad\qquad
+[A(t_3),[A(t_2),A(t_1)]]
\Big).
\end{align*}
For the single-harmonic forcing considered here, evaluation of these
integrals over $T=2\pi/\omega$ gives
\begin{equation*}
\Omega_3=
\frac{2\pi\varepsilon}{\omega^3}
[A_0,[A_0,H]]
-\frac{\pi\varepsilon^2}{2\omega^3}
[H,[A_0,H]].
\end{equation*}

Therefore,
\[
\begin{aligned}
A_{\mathrm{eff}}
&=
\frac{\Omega(T)}{T}
=
A_0
+
\frac{\epsilon}{\omega^2}
[A_0,[A_0,H]] \\
&\quad
-
\frac{\epsilon^2}{4\omega^2}
[H,[A_0,H]]
+
O(\omega^{-3}) .
\end{aligned}
\]
and hence, for fixed $\epsilon$,
\[
A_{\mathrm{eff}}
=
A_0+O(\omega^{-2}),
\qquad
\omega\rightarrow\infty.
\]

Provided the dominant Floquet exponent remains isolated and depends
smoothly on the high-frequency perturbation, it follows that
\[
\Lambda(\omega)\rightarrow
\Lambda_0
=
\max_j\operatorname{Re}\lambda_j(A_0)
\qquad
(\omega\rightarrow\infty).
\]

 }

Comparison between the numerical and analytical stability boundaries demonstrates that the theoretical approximation captures the overall geometry of the phase diagrams and correctly predicts the direction of the stability shifts induced by periodic forcing in the adiabatic limit ($\omega \ll 1$). While quantitative deviations become apparent in regions of strong forcing and near critical transitions, the analytical theory successfully reproduces the principal features of the stability landscape across a broad range of ecological parameters. Taken together, the phase diagrams reveal that periodic forcing does not merely perturb ecological communities but fundamentally reorganizes their stability boundaries. Most notably, increasing the forcing frequency systematically enlarges the stable region of parameter space, indicating that the timescale of environmental fluctuations plays a central role in determining ecosystem resilience.

To compare the influence of interaction topology on ecosystem resilience, we evaluate the stability phase diagrams for predator-prey and competition-mutualism architectures across a wide range of forcing frequencies in Fig.~\ref{fig:topological_comparison}. The maximal Lyapunov exponent $\Lambda$ is shown as a function of the forcing frequency $\omega$ and either the interaction strength $\sigma$ (panels a,c) or the self-regulation rate $d$ (panels b,d). The numerical results are consistent with the stability hierarchy predicted by the correlation coefficient $\tau$ in the adiabatic limit ($\omega \ll 1$). Comparing the top and bottom rows reveals a pronounced topological contrast: predator-prey communities ($\tau<0$; panels a,b) possess a substantially larger stable region ($\Lambda<0$) than competition-mutualism community ($\tau>0$; panels c,d). For example, in the low-frequency regime, the critical interaction strength at which $\Lambda$ changes sign is substantially larger for predator-prey communities than for competition-mutualism community, demonstrating the enhanced robustness of predator-prey structured communities. These results support the analytical prediction that antisymmetric predator-prey interactions suppress the spectral edge, while symmetric competitive and mutualistic interactions increase it, thereby reducing community stability.

Figure~\eqref{fig:topological_comparison} also highlights the range of validity of the adiabatic approximation. In the low-frequency regime ($\omega\ll1$), the numerically determined stability boundary (solid black line) exhibits excellent agreement with the analytical prediction given by Eq.~(\ref{eq:lambda_analytical}) (dashed line). In this regime, the community effectively tracks the slowly varying environment, and its stability is governed by the time-averaged rightmost spectral edge of the instantaneous community matrix. As the forcing frequency increases, systematic deviations from the adiabatic prediction emerge. The numerical stability boundary bends away from the frequency-independent theoretical result and progressively expands the stable parameter region. Consequently, interaction strengths that would destabilize the system under slow environmental forcing can remain stable when the forcing varies rapidly (Fig.~\ref{fig:topological_comparison}a,c). Similarly, the minimum self-regulation strength required for stability decreases substantially at high frequencies (Fig.~\ref{fig:topological_comparison}b,d). {\color{black}These results demonstrate a frequency-induced stabilization effect that lies beyond the adiabatic approximation and can be interpreted in the high-frequency regime using dynamical averaging and the Magnus expansion.}


\subsection{Validation of the Analytical Approximation}

To assess the validity of the analytical approximation (Eq.~\ref{eq:lambda_analytical}), we perform a systematic statistical comparison between the numerically computed maximal Lyapunov exponent $\Lambda_{\rm num}$ and the theoretical prediction $\Lambda_{\rm th}$ for all three network architectures. For each pair $(C,\omega)$, with $C\in\{0.1,0.2,0.4,0.6,0.8,1.0\}$ and $\omega\in\{10^{-2},10^{-1},10^0,10^1\}$, we vary six evenly spaced values of $\sigma\in[0.3,1.5]$, $d\in[0,1.5]$, and $\varepsilon\in[0,4]$, yielding $216$ parameter combinations per grid point. For every value of $\sigma$, a single realization of the matrices $B$ and $H$ is generated and subsequently held fixed while varying $d$ and $\varepsilon$. New independent realizations are generated whenever $C$, $\omega$, or $\sigma$ changes. For each parameter combination, $\Lambda_{\rm num}$ and $\Lambda_{\rm th}$ are computed and compared through a linear regression \cite{montgomery2021introduction,press2007numerical} $\Lambda_{\rm num}=m\Lambda_{\rm th}+c$. The fit slope $m$, intercept $c$, coefficient of determination $R^2$, and mean absolute error are used to quantify agreement between theory and simulation.

Figures~\eqref{fig:random_validation}, \eqref{fig:predatorprey_validation}, and \eqref{fig:compmut_validation} summarize the validation results for random, predator-prey, and competition-mutualism community, respectively. A striking result is the robustness of the analytical approximation across all three topological classes. For forcing frequencies $\omega\le10^{-1}$, the regression slopes remain
close to unity, the intercepts remain small, and the coefficients of
determination typically exceed $0.95$ throughout most of the connectance
range. The mean absolute error is simultaneously small, demonstrating
that the adiabatic approximation accurately captures the numerical
dynamics of slowly varying environments irrespective of the underlying
interaction topology.

Importantly, the quality of the approximation depends far more strongly on the forcing frequency than on the network connectance. Across all three architectures, the validation metrics exhibit only a weak dependence on $C$, whereas a pronounced deterioration occurs as $\omega$ increases. This observation supports the central assumption underlying the theory: the dominant source of error is the breakdown of the adiabatic approximation rather than inaccuracies in the random-matrix estimate of the instantaneous spectral edge. As the forcing frequency enters the non-adiabatic regime ($\omega\gtrsim1$), the fit slopes systematically decrease below unity, the coefficients of determination deteriorate, and the mean absolute errors increase substantially. {\color{black}The effect is strongest at $\omega=10$ (the largest deviation observed within the range numerically considered), where the environmental modulation evolves on timescales comparable to or faster than the intrinsic community relaxation time. Note that, here $\omega=10$ is simply the largest frequency examined numerically; the high-frequency analysis concerns the asymptotic limit $\omega\rightarrow\infty$.}


Although the three network architectures possess markedly different
stability properties through their distinct correlation coefficients
$\tau$, the regime of validity of the analytical approximation remains
qualitatively unchanged. predator-prey communities
($\tau=-2/\pi$) are the most stable, competition-mutualism community
($\tau=2/\pi$) are the least stable, and random communities
($\tau=0$) occupy an intermediate position, in agreement with the
ordering predicted by Eq.~(\ref{eq:lambda_analytical}).
Nevertheless, all three architectures display the same transition from
excellent agreement in the adiabatic regime to systematic deviations at
high forcing frequencies. This demonstrates that the breakdown of the
theory is governed primarily by the temporal dynamics of the forcing,
rather than by the detailed topology of the ecological network.

\begin{figure*}[t]
    \centering
    \includegraphics[width=\textwidth]{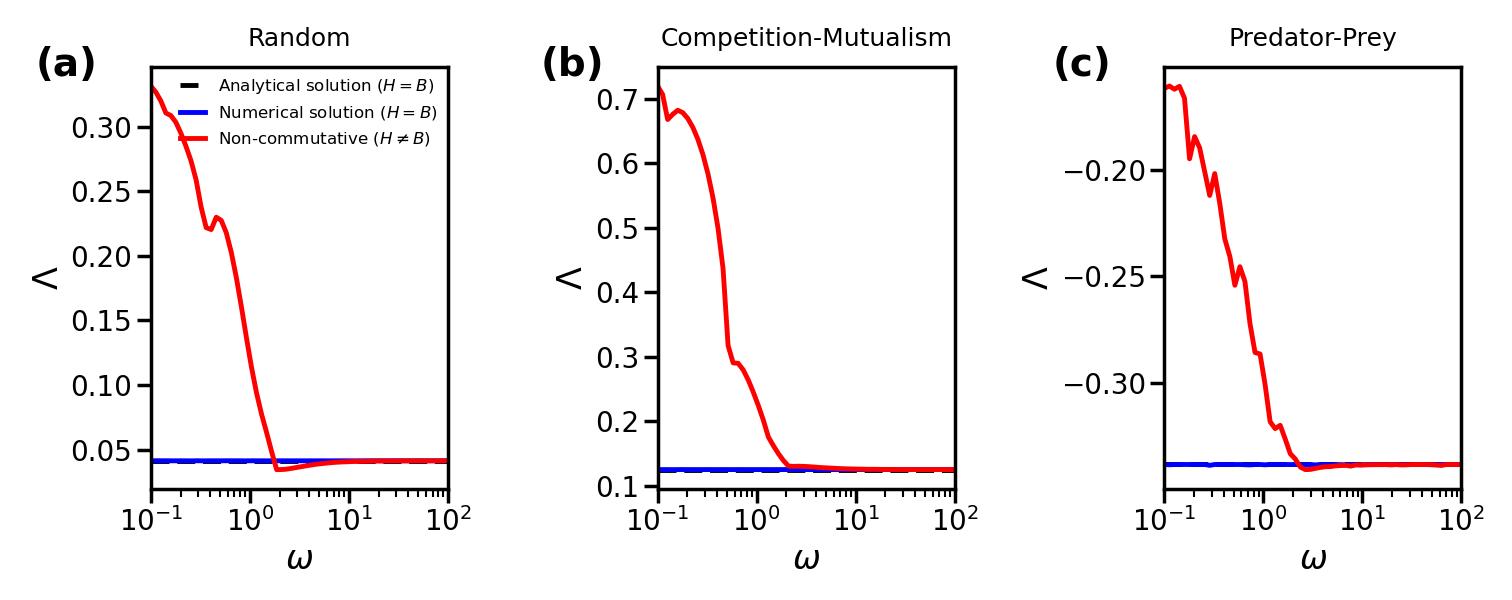} 
    \caption{{\bf Comparison of the $H=B$ and $H\neq B$ periodically forced ecological communities.} The maximal Lyapunov exponent $\Lambda$ is illustrated as a function of the forcing frequency $\omega$ for (a) random, (b) competition-mutualism, and (c) predator-prey interaction networks. The black dashed line indicates the analytical prediction $\Lambda = -d + \lambda_{\max}(B)$ (Eq.\ \eqref{eq:lyapunov_exponent}), which remains invariant under temporal forcing when the perturbation matrix $H$ is identical with the interaction matrix $B$ ($H=B$, blue solid line). The numerical estimates for the commutative case $H=B$ (blue) exhibit high precision, showing persistent agreement with the analytical prediction across the entire frequency spectrum. In contrast, the general non-commutative case ($H \neq B$, red solid line) reveals a complex frequency-dependent stability profile. Specifically, low-frequency forcing induces a destabilizing effect, whereas high-frequency forcing leads to an emergent rescue effect, where $\Lambda$ asymptotically converges to the system's intrinsic stability limit as $\omega \to \infty$. Notably, the stability hierarchy is preserved across all architectures: predator-prey communities exhibit the most negative (stabilizing) Lyapunov exponents, random communities occupy an intermediate regime, and competition-mutualism communities manifest the largest (least stable) Lyapunov exponents. Simulation parameters: community size $S=100$, forcing amplitude $\varepsilon=2.0$, connectance $C=0.4$, interaction strength $\sigma=0.8$, and self-regulation rate $d=0.5$.}
    \label{fig:Figure_8}
\end{figure*}

\section{Discussion}

Predicting the stability of complex ecosystems in fluctuating environments remains one of the most formidable challenges in theoretical ecology. Traditionally, investigating the resilience of large, non-autonomous ecological networks has required computationally expensive numerical integration. Furthermore, the stability of ecological communities emerges from a complex interplay between interaction topology, interaction strength, environmental variability, and intrinsic regulatory mechanisms. By combining random matrix theory with non-autonomous dynamical systems, the present study provides a unified minimal model for understanding how these factors jointly determine ecosystem stability under periodic environmental forcing. We successfully bridge Random Matrix Theory with direct numerical computation of maximal Lyapunov exponents to develop a rigorous analytical framework that predicts the stability of periodically forced ecological networks.

A major outcome is the identification of a simple analytical approximation for the maximal Lyapunov exponent in periodically forced ecological systems. Under the assumptions of a large system size and slow environmental forcing, we demonstrate that the stability of the time-varying ecosystem under consideration can be approximated analytically using simplified assumptions. Within the adiabatic regime, the Lyapunov exponent is determined by the time-averaged rightmost spectral edge of the instantaneous community matrix. The resulting expression \eqref{eq:lambda_analytical}
successfully predicts both the ordering of network architectures and the locations of numerical stability boundaries. Extensive statistical validation across broad regions of parameter space demonstrates that the approximation remains highly accurate under the adiabatic approximation.

Another key finding of this study is the existence of a robust topological stability hierarchy. Across all parameter regimes considered, predator-prey communities consistently exhibit the lowest maximal Lyapunov exponents, competition-mutualism communities display the highest values, and random communities occupy an intermediate position. This ordering is elegantly encapsulated by the structural correlation coefficient $\tau$, which enters the analytical approximation through the elliptic law. Negative correlations in reciprocal interactions reduce the spectral edge, thereby enhancing stability, whereas positive correlations enlarge the spectral edge, promoting instability. These findings are consistent with long-standing theoretical arguments \cite{allesina2012stability,mougi2012diversity,allesina2008network} suggesting that antisymmetric predator-prey interactions ($\tau < 0$) generate stabilizing feedback loops while symmetric competition-mutualism interactions ($\tau > 0$) create positive feedback mechanisms that amplify perturbations. Random communities ($\tau = 0$) serve as the structural baseline between these two extremes. Crucially, in the adiabatic limit ($\omega \ll 1$), our analytical boundary (Eq.~\eqref{eq:lambda_analytical}) exhibits near-perfect predictive accuracy, demonstrating that slowly changing environments force the ecosystem to continuously track its instantaneous leading eigenvalue.

However, the most profound insight of this study emerges from the systematic breakdown of the adiabatic approximation at higher forcing frequencies. As the forcing frequency increases, the exact stability boundaries systematically deviate from the adiabatic prediction and shift toward larger values of interaction strength and forcing amplitude.  Rather than destabilizing the ecosystem, this high-frequency forcing induces a positive rescue effect, systematically expanding the stable parameter regime. 
The dynamical averaging prevents perturbations from fully exploiting the instantaneously unstable directions of the phase space. This effect is particularly evident in the $(\varepsilon,\omega)$ phase diagram, where increasingly rapid environmental oscillations allow communities to tolerate progressively stronger fluctuations. When environmental forcing varies slowly, perturbations experience sustained intervals of enhanced interaction strengths, allowing deviations from equilibrium to accumulate. In contrast, rapid oscillations alternate between favorable and unfavorable conditions on timescales shorter than the community can effectively respond. Positive and negative perturbative contributions therefore cancel over time, reducing their net influence on long-term growth rates. In the asymptotic limit of very large forcing frequency, the oscillatory contribution becomes increasingly ineffective, and the dynamics approach those of the unforced system (See Fig.\ \eqref{fig:Figure_8}). The observed high-frequency rescue effect is therefore analogous to averaging phenomena that arise in periodically driven physical and biological systems.

{\color{black}The Magnus expansion also provides a useful interpretation of the intermediate-frequency regime. For the present single-frequency forcing, the leading correction to the averaged generator scales as $O(\omega^{-2})$ and depends on nested commutators between the unforced interaction matrix and the environmental susceptibility matrix. Thus, as $\omega$ decreases from the high-frequency regime, these corrections become progressively more important, providing a
mechanism for the departure from the high-frequency limiting
behavior observed numerically. At intermediate frequencies, however, higher-order Magnus terms may become non-negligible, so the truncated high-frequency expansion should not be regarded as a quantitatively controlled description throughout the entire crossover regime.}


The present framework also clarifies the distinct roles played by ecological parameters. Increasing interaction strength $\sigma$, connectance $C$, or forcing amplitude $\varepsilon$ generally promotes instability by enlarging the effective spectral radius of the interaction matrix. {\color{black}In contrast, stronger self-regulation $d$ shifts the spectrum toward negative values and therefore stabilizes the local perturbation dynamics.} Interestingly, the dependence on the community size $S$ weakens once the system is sufficiently large, indicating convergence toward the large-$S$ random-matrix limit assumed by the theory.


While the present framework provides a useful theoretical foundation for understanding the stability of periodically forced ecological networks, several important extensions remain to be explored. First, our model is based on a linearized description of community dynamics near equilibrium. Incorporating nonlinear ecological processes, such as saturating functional responses, density-dependent regulation, or species-specific carrying capacities, would enable the investigation of ecosystem resilience far from equilibrium and may reveal transitions to complex dynamical states, including limit cycles and chaos.

Moreover, environmental variability in natural ecosystems is rarely purely periodic. Extending the present framework to account for stochastic environmental fluctuations would provide a more realistic description of ecological systems subject to unpredictable climatic and environmental perturbations. Such an approach would allow investigation of the interplay between deterministic seasonal forcing and random environmental noise, and how these jointly influence community stability and resilience. {\color{black}These limitations define the scope of the present framework. The considered model provides an analytical description of local perturbation stability under periodic environmental modulation, while extensions to nonlinear, species-specific, stochastic, or mechanistically driven ecological dynamics with appropriate empirical data remain important directions for future work.}



Overall, the results demonstrate that ecological stability under environmental forcing is governed jointly by interaction topology and forcing timescale. While network structure determines a community's baseline susceptibility to instability, the temporal characteristics of environmental fluctuations can dramatically alter stability outcomes. The combination of random matrix theory, Lyapunov exponent analysis, and periodically driven ecological dynamics, therefore, offers a powerful framework for understanding ecosystem resilience in fluctuating environments.

\section{Conclusion}

In this study, we have investigated the stability of periodically forced ecological communities using a combination of Random Matrix Theory and Lyapunov exponent analysis. By considering three different interaction architectures subjected to time-dependent environmental forcing, we have developed an analytical framework that connects ecological interaction topology, environmental variability, and dynamical stability through the maximal Lyapunov exponent.


In the adiabatic limit, where environmental changes are much slower than the system’s internal dynamics, we have derived an analytical approximation of the maximal Lyapunov exponent by time-averaging the largest eigenvalue of the interaction matrix. This result separates the influence of interaction topology, captured by the structural correlation coefficient $\tau$, from environmental forcing, described by the amplitude-dependent factor $M(\varepsilon)$, which measures the impact of fluctuation amplitudes. The resulting theory has accurately predicted stability boundaries across a broad region of parameter space and has correctly reproduced the observed topological stability hierarchy, with predator-prey communities being the most stable, competition-mutualism communities the least stable, and random communities occupying an intermediate position. Extensive numerical validation has further confirmed the accuracy of the approximation in the low-frequency regime.

Beyond the adiabatic limit, our numerical investigations have revealed a pronounced high-frequency stabilization effect. We have shown that rapid environmental oscillations systematically suppress the maximal Lyapunov exponent and expand the stable region of parameter space, thereby stabilizing ecosystems. This high-frequency rescue effect highlights the crucial role of environmental timescales and demonstrates that the temporal characteristics of environmental variability can be as important as interaction topology in determining ecosystem resilience.

Overall, our results have established a unified framework for understanding the stability of complex ecological networks under periodic environmental forcing. By integrating Random Matrix Theory with explicit time-dependent dynamics, we have provided a predictive approach to quantifying ecosystem resilience in fluctuating environments and a foundation for future studies that incorporate nonlinear interactions, stochastic forcing, and more realistic ecological complexity. {\color{black}Beyond its ecological motivation, the framework developed here provides a general mathematical perspective for studying stability in high-dimensional nonautonomous networks subject to time-dependent forcing.}

\appendix
\section{The Special Case $H=B$} \label{Appendix}


For completeness, we briefly discuss the special limiting case $H= B$, although this scenario is not the primary focus of the present work. Ecologically, this assumption implies that every ecological interaction responds identically to environmental forcing. In other words, the environment uniformly amplifies or weakens all interactions by the same multiplicative factor, preserving not only the interaction topology but also the relative strengths of all species interactions. Such perfectly synchronized environmental responses are generally unrealistic in natural ecosystems, where different interactions are expected to exhibit distinct sensitivities to environmental variability. Nevertheless, this special case is mathematically instructive because it admits an exact analytical solution.

\subsection{Preservation of the interaction correlation}

When $H=B$, the instantaneous interaction matrix becomes
\begin{equation}
A(t) = -dI+\left[1+\varepsilon\cos(\omega t)\right]B.
\label{eq:appendix_A}
\end{equation}
Since every matrix element is multiplied by the same scalar factor,
\[
A_{ij}(t) = \left[1+\varepsilon\cos(\omega t)\right]B_{ij},
\]
for $i \neq j$, the covariance between reciprocal interactions satisfies
\[
\mathrm{Cov}(A_{ij},A_{ji}) = \left[1+\varepsilon\cos(\omega t)\right]^2 \mathrm{Cov}(B_{ij},B_{ji}),
\] 
while
\[
\mathrm{Var}(A_{ij}) = \left[1+\varepsilon\cos(\omega t)\right]^2 \mathrm{Var}(B_{ij}).
\]
Therefore, the reciprocal interaction correlation coefficient of the instantaneous matrix is
\[
\tau_A = \frac{\mathrm{Cov}(A_{ij},A_{ji})}{\sqrt{\mathrm{Var}(A_{ij}) \mathrm{Var}(A_{ji})}} = \tau_B.
\]
Thus, the environmental forcing preserves the correlation structure of the ecological network exactly. 

\subsection{Adiabatic approximation}

Since the reciprocal interaction correlation remains unchanged, the elliptic-law approximation immediately yields the instantaneous spectral edge as
\[
\lambda_{\max}(t) = -d + (1+\tau)\sigma\sqrt{C} \left|1+\varepsilon\cos(\omega t)\right|,
\]
where the absolute value arises because the interaction variance is multiplied by
\[
\left(1+\varepsilon\cos(\omega t)\right)^2.
\]
Applying the same adiabatic argument developed in the main text gives
\[
\Lambda = -d + (1+\tau)\sigma\sqrt{C} \frac{1}{2\pi} \int_0^{2\pi} \left|1+\varepsilon\cos\theta\right| \, d\theta,
\]
which constitutes the corresponding adiabatic approximation for this special case.

For forcing amplitudes $0 \le \varepsilon \le 1$, the environment does not invert the interaction signs of the ecosystem. Because $1 + \varepsilon \cos\theta \ge 1 - \varepsilon \ge 0$, the absolute value in the adiabatic integral is redundant. The time-averaging factor evaluates exactly to unity:
$$M_B(\varepsilon) = \frac{1}{2\pi} \int_0^{2\pi} (1 + \varepsilon \cos\theta) \, d\theta = 1.$$

Interestingly, this proves that the forcing amplitude $\varepsilon$ has no effect whatsoever on the adiabatic stability prediction in this regime for $B=H$.

\subsection{Exact solution}

Unlike the general model considered in the main text, Eq.\ \eqref{eq:appendix_A} possesses an important algebraic property. Since every instantaneous matrix is proportional to the same constant matrix $B$,
\[
A(t_1) = -dI+f(t_1)B, \qquad A(t_2) = -dI+f(t_2)B,
\]
where
\[
f(t)=1+\varepsilon\cos(\omega t),
\]
it immediately follows that
\[
[A(t_1),A(t_2)] = 0
\]
for all pairs of times $t_1$ and $t_2$. Consequently, the time-ordered exponential normally required for non-autonomous systems reduces to an ordinary matrix exponential,
\[
\mathbf{x}(t) = \exp \left[ \int_0^t A(s) \, ds \right] \mathbf{x}(0).
\]
Performing the integration gives
\[
\int_0^t A(s) \, ds = -dt I + \left( t+ \frac{\varepsilon}{\omega} \sin(\omega t) \right)B.
\]
Hence,
\[
\mathbf{x}(t) = e^{-dt} \exp \left[ \left( t+ \frac{\varepsilon}{\omega} \sin(\omega t) \right)B \right] \mathbf{x}(0).
\]
Let $\lambda_{\max}(B)$ denote the rightmost eigenvalue of the baseline interaction matrix. The asymptotic growth of the perturbation therefore satisfies
\[
\|\mathbf{x}(t)\| \sim \exp \left[ \left( -d+\lambda_{\max}(B) \right)t + \frac{\varepsilon}{\omega} \lambda_{\max}(B) \sin(\omega t) \right].
\]
The maximal Lyapunov exponent is then
\begin{equation}
\Lambda = \lim_{t\rightarrow\infty} \frac{1}{t} \ln \|\mathbf{x}(t)\| = -d+\lambda_{\max}(B),
\label{eq:lyapunov_exponent}
\end{equation}
because
\[
\lim_{t\rightarrow\infty} \frac{\sin(\omega t)}{t} = 0.
\]
Remarkably, this result is exact and holds for arbitrary forcing amplitude $\varepsilon$ and forcing frequency $\omega$. Thus, unlike the general model studied throughout this work, no adiabatic approximation is required. The exact analytical result \eqref{eq:lyapunov_exponent} is validated by direct numerical integration, as shown in Fig.~\eqref{fig:Figure_8}. Across all three ecological communities, the numerical maximal Lyapunov exponents for the commutative case $H=B$ are in excellent agreement with the theoretical prediction, confirming the validity of the derivation. 

This exact solvability highlights an important mathematical distinction between the two formulations. When $H=B$, all instantaneous interaction matrices commute, causing the periodic environmental force to influence only the transient dynamics while leaving the asymptotic exponential growth rate unchanged. In contrast, the general model employs two independent interaction matrices, for which the instantaneous matrices generally do not commute, i.\ e.\ , $[B, H] \neq 0$. This non-commutativity prevents exact integration, necessitates the adiabatic approximation in the low-frequency regime, and ultimately gives rise to the rich frequency-dependent stability behavior and high-frequency stabilization reported in the main text.

\section*{AUTHORS’ CONTRIBUTIONS}

 {\bf Sayantan Nag Chowdhury:}  Conceptualization, Methodology, Software, Validation, Formal analysis, Investigation, Visualization, Writing - original draft, Writing - review \& editing.

\section*{Competing interests.} We declare we have no competing interests.

\section*{ACKNOWLEDGMENTS}

I would like to thank my wife, Dr.\ Srilena Kundu (Helmholtz Institute for Functional Marine Biodiversity at the University of Oldenburg [HIFMB] and the Alfred-Wegener Institute, Helmholtz Center for Polar and Marine Research), for her insightful discussions and careful reading of the manuscript.

    

\section*{DATA AVAILABILITY}
The code supporting the findings of this study are openly available on Github at \cite{web_1}.

\bibliographystyle{apsrev4-1}
\bibliography{Sayantan}

\end{document}